\documentclass{article}
\usepackage{float}
\usepackage[a4paper, total={7in, 9in}]{geometry}

\usepackage{amsmath,amssymb,amsthm,mathtools,bm,booktabs,graphicx,array,multirow}
\usepackage{natbib}
\usepackage{xcolor}
\usepackage[hidelinks]{hyperref}
\usepackage{url}
\usepackage{subfig} 
\newtheorem{theorem}{Theorem}
\newtheorem{proposition}[theorem]{Proposition}
\newtheorem{corollary}[theorem]{Corollary}

\theoremstyle{remark}
\usepackage{subcaption}

\newtheorem*{remark}{Remark}
\newcommand{\T}{\mathbb T}
\newcommand{\R}{\mathbb R}

\newcommand{\dd}{\,\mathrm d}

\newcommand{\E}{\mathbb E}

\usepackage{algorithm}
\usepackage{algpseudocode}

\newcommand{\Z}{\mathbb{Z}}
\newcommand{\DE}{D_{\!E}}
\newcommand{\DEpar}[1]{D_{\!E,#1}}
\newcommand{\dg}{d_g}
\newcommand{\wrap}{\operatorname{wrap}}
\newcommand{\ARI}{\operatorname{ARI}}

\newcommand{\footremember}[2]{%
	\footnote{#2}
	\newcounter{#1}
	\setcounter{#1}{\value{footnote}}%
}
\newcommand{\footrecall}[1]{%
	\footnotemark[\value{#1}]%
} 
\title{How to bowl properly}
\author{%
	The Dude\footremember{alley}{Holly Star Lanes Bowling Alley}%
	\and Walter Sobchak\footremember{trailer}{Probably in a trailer park}%
	\and Jesus Quintana\footrecall{alley} \footnote{Mexico?}%
	\and Uli Kunkel\footrecall{trailer} \footnote{Germany?}%
}

\begin{document}
	
	

	\title{Geometry-Adaptive Clustering on High-Dimensional Tori}

\author{ Surojit Biswas\thanks{Email: sbiswas1@uni-bonn.de} \footremember{Bonn}{Hausdorff Research Institute for Mathematics, University of Bonn} \and Soumyadip Paik \footremember{Amex}{American Express} \and Buddhananda Banerjee \footremember{IITKGP}{Department of Mathematics, Indian Institute of Technology Kharagpur} }
	
	\date{}
	
	\maketitle
\begin{abstract}
Multivariate angular data lie on a torus, but clustering methods often
use the flat torus. We introduce curved-torus energy (CTE), a
position-dependent dissimilarity based on restricted Riemannian path
energy and minimization over winding classes. CTE has a certified
finite winding reduction, an exact flat endpoint, a geodesic upper
bound, and stability with respect to the geometry parameter. For
label-free geometry selection, we propose CEHS-HM, a held-out
criterion for selecting the warping parameter at a specified
clustering resolution. We establish a finite-grid population-score
regret bound and also consider joint geometry-resolution selection
over finite candidate grids. Controlled experiments on
$\mathbb{T}^3$ show that CTE agrees with the flat method at zero
warping and becomes more useful as geometric warping increases.
Additional experiments show that the CEHS-selected geometry should be
viewed as a data-adaptive clustering choice rather than as an
estimator of a generative deformation parameter. Under a
product-von-Mises mechanism, the selector tends toward the flat
endpoint when curvature is unnecessary. On ILE protein torsion data,
the proposed method identifies a stable five-mode representation of
joint backbone-side-chain conformation.
\end{abstract}


\section{Introduction}
Vectors of angles arise in many applications, including  ophthalmology \cite[see][]{banerjee2026intrinsic, jha2018circular},  robotics \cite[see][]{mescher2024geometric}, meteorology \cite[see][]{jha2017multiple, biswas2024intrinsic}, molecular
conformations angles \cite[see][]{Eltzner2018torus}, biomechanics \citealp[see][]{sato2023online}, and directional time series \citep{giannakis2019data}. When several
angular variables are observed jointly, their natural sample space is
the product torus
$\T^p=(S^1)^p$, rather than $\R^p$.

Directional statistics provides probability models and inferential
tools for circular, spherical, and toroidal data
\citep{biswas2026efficient,mardiajupp2000,pewseygarcia2021,mardia2007,mardia2012}.
The conventional statistical representation of multivariate angular
data is the flat product torus $(\mathbb{S}^1)^p$, with dependence
typically introduced through an angular density. Correspondingly,
existing toroidal clustering methods have been developed using density
estimation, mixture models, dimension reduction, and
principal-submanifold representations
\citep{jung2021clustering,hongjung2022,mardia2022pca,wuyao2026}. A complementary question is less often treated as part of the statistical problem itself: \emph{which geometry should define similarity between toroidal observations?}

The usual product-torus metric assigns the same local weight to an
angular displacement regardless of position. By contrast, on the
curved torus in \eqref{eq:torus-parametrization}, with radii
$R>r>0$,
$$
\dd s^2
=
r^2\dd\theta_1^2
+
(R+r\cos\theta_1)^2\dd\theta_2^2
$$ is  the induced Riemannian metric, see \eqref{eq:classical-torus-metric}.
Hence the metric weight associated with displacement in the
$\theta_2$ direction depends on $\theta_1$, the same angular
increment, which $\dd\theta_2$ produces different Riemannian lengths at
different positions on the torus.
This motivates the coordinatewise-standardized statistical metric
$$
\dd\widetilde s_\lambda^2
=
\dd\theta_1^2
+
(1+\lambda\cos\theta_1)^2\dd\theta_2^2,
$$
where $0 < \lambda<1,$ and  with $\lambda=0$ recovering the flat product torus. Importantly, this standardization divides the two orthogonal scale factors by their own baseline radii; it is not division of the original induced metric by a single global constant.

Riemannian clustering methods typically regard geometry as fixed in advance \citep{pennec2006,subbarao2009,wang2013}. Metric-learning work instead adapts distances or representations from data, including extensions to manifolds \citep{aaltoverma2019} and unsupervised metric learning \citep{dutta2020unsupervised}. Our setting is narrower and more structured, rather than learning an unrestricted metric, we select from an interpretable family of position-dependent toroidal Riemannian geometries with an exact flat member. The selection criterion is also related to cluster-stability and held-out validation methods such as stability-based model selection \citep{benhur2002stability}, Prediction Strength \citep{tibshiraniwalther2005}, and the Gap statistic \citep{tibshirani2001gap}; here the held-out objective is used primarily to tune geometry for a specified clustering resolution.

We start from the recursive embedding of $\T^N$ into $\R^{N+1}$ used by \citet{llibre2024}; the embedding itself, which is not our contribution. Our contributions are\textit{ \textbf{(1)} a dimensionless, position-dependent recursive torus geometry with an exact flat endpoint;
\textbf{(2)} CTE, a winding-aware dissimilarity with a certified finite search over winding classes and explicit comparison with intrinsic geodesic and flat-torus distances;
\textbf{(3)} CEHS-HM, a label-free held-out criterion for selecting the clustering geometry at fixed $K$, together with finite-sample score properties and a finite-grid population-score regret guarantee; and
\textbf{(4)} controlled and misspecified simulations that separate geometry adaptation from generative-parameter recovery, plus a biological application to joint protein torsion angles.}

% \begin{enumerate}
% \item a dimensionless, position-dependent recursive torus geometry with an exact flat endpoint;
% \item CTE, a winding-aware dissimilarity with a certified finite search over winding classes and explicit comparison with intrinsic geodesic and flat-torus distances;
% \item CEHS-HM, a label-free held-out criterion for selecting the clustering geometry at fixed $K$, together with finite-sample score properties and a finite-grid population-score regret guarantee; and
% \item controlled and misspecified simulations that separate geometry adaptation from generative-parameter recovery, plus a biological application to joint protein torsion angles.
% \end{enumerate}
% Detailed proofs, secondary model-selection experiments, full $\T^2$ benchmark diagnostics, scalability results, and extended geodesic checks are retained in the Supplementary Material.

\section{Curved-torus energy dissimilarity}
\label{sec:cte}  
\subsection{Recursive standardized geometry}
\label{sec:recursive-geometry}
For the ordinary two-dimensional curved torus with angular coordinates
$(\theta_1,\theta_2)$ and radii $R>r>0$,
\begin{eqnarray}
	\label{eq:torus-parametrization}
	\mathbf{X}(\theta_1,\theta_2)
	& = & \Big[(R+r\cos\theta_1)\cos\theta_2,(R+r\cos\theta_1)\sin\theta_2, 
	r\sin\theta_1 \Big]
\end{eqnarray}
$	\theta_1,\theta_2\in[0,2\pi),$
the induced Riemannian metric is
\begin{equation}
	\label{eq:classical-torus-metric}
	\dd s^2
	=
	r^2\,\dd\theta_1^2
	+
	(R+r\cos\theta_1)^2\,\dd\theta_2^2.
\end{equation}
Thus, even for the classical curved torus, the cost of displacement in the
second angular direction depends on the current value of the first angle.

Equivalently, the metric may be written in terms of its orthogonal scale
factors as
$\dd s^2
=
h_1^2\,\dd\theta_1^2
+
h_2^2\,\dd\theta_2^2,
\quad
h_1=r,
\quad
h_2=R+r\cos\theta_1.$

To remove the dimensional baseline radii while retaining the
position-dependent deformation, we normalize these scale factors by
their natural reference scales, $\widetilde h_1
=
\frac{h_1}{r}
=
1,
\quad
\widetilde h_2
=
\frac{h_2}{R}
=
1+\frac{r}{R}\cos\theta_1.$
Writing $\lambda=\frac{r}{R},
\quad 0<\lambda<1,$
motivates the dimensionless one-parameter metric family
\begin{equation}
	\label{eq:t2metric}
	\dd\widetilde s_\lambda^2
	=
	\dd\theta_1^2
	+
	(1+\lambda\cos\theta_1)^2\,\dd\theta_2^2,
	\quad
	0\leq\lambda<1.
\end{equation}
For a physical curved torus $\lambda=r/R$; in the statistical family, $\lambda$ is a \emph{geometry} or \emph{warping} parameter. Because the coordinate scale factors are standardized individually, \eqref{eq:t2metric} should be viewed as a statistical construction inspired by the induced metric, not as the induced metric divided by a global radius scale.  At this endpoint $\lambda=0$, $\dd\widetilde s_0^2
=
\dd\theta_1^2+\dd\theta_2^2,$
so the standard flat product-torus metric is recovered exactly.  

To extend this position-dependent geometry to higher-dimensional tori,
we start from the recursive construction of \citet{llibre2024}. For $[\theta]=(\theta_1,\theta_2,\cdots, \theta_N) \in\T^N,$ its
radii satisfy
$\rho_1=R_1,
\quad
\rho_k
=
R_k+\rho_{k-1}\cos\theta_{k-1},
\quad
k\geq2.$
The corresponding embedding may be written as
\[
\begin{aligned}
	F_N(\theta)
	=
	\bigl(
	&\rho_N\cos\theta_N,\,
	\rho_N\sin\theta_N,\,
	\rho_{N-1}\sin\theta_{N-1},\\
	&~~~~~~~~~~~~~~~~~~~\ldots,\,
	\rho_1\sin\theta_1
	\bigr).
\end{aligned}
\]
Its Euclidean induced metric is diagonal. Indeed, the recursive
identity
$\dd(\rho_k\cos\theta_k)^2
+
\dd(\rho_k\sin\theta_k)^2
=
\dd\rho_k^2
+
\rho_k^2\dd\theta_k^2$
eliminates the cross terms successively, yielding  $	g
=
\operatorname{diag}
\bigl(
\rho_1^2,\ldots,\rho_N^2
\bigr).$
%	For statistical modeling, we remove the arbitrary baseline radius
%	scales while retaining the position-dependent modulation of the
%	angular directions	
Now	 under a common dimensionless radius ratio
$\lambda$, for $
		k=2,\ldots,N,$ this leads to the normalized recursive family
\begin{equation}
	\label{eq:hk}
	\begin{aligned}
		\dd\widetilde s_\lambda^2
		&=
		\sum_{k=1}^{N}
		h_k(\theta;\lambda)^2
		\dd\theta_k^2, ~~~~h_1(\theta;\lambda)
		=1,
		\\
		h_k(\theta;\lambda)
		&=
		1+\lambda
		h_{k-1}(\theta;\lambda)
		\cos\theta_{k-1}.
	\end{aligned}
\end{equation}

The normalization from the underlying recursive
scale factors and the corresponding multi-parameter form are given in
Appendix~\ref{app:metric-normalization}. 
It is therefore a	coordinatewise baseline standardization of the induced scale factors, introduced as a statistical modelling construction while preserving the
position-dependent anisotropy inherited from the recursive embedding.

For \eqref{eq:hk}, the corresponding diagonal Riemannian metric is
$$\displaystyle
g_\lambda(\theta)
=
\operatorname{diag}\!\left(
h_1(\theta;\lambda)^2,\ldots,
h_N(\theta;\lambda)^2
\right).
$$
Thus, for a tangent vector
$v=(v_1,\ldots,v_N)\in T_\theta\T^N$,
$
g_{\lambda,\theta}(v,v)
=
\sum_{k=1}^N
h_k(\theta;\lambda)^2v_k^2,
$ and 
the metric
coefficients must remain strictly positive. Moreover, because
$\lambda$ will subsequently be treated as a geometry-selection
parameter, it is useful to establish positivity uniformly over an
admissible range of $\lambda$. The following result provides both a
global sufficient condition and uniform nondegeneracy of the resulting
metric family.

%For the curved torus
%\begin{equation}
%\label{eq:torus-parametrization}
%\mathbf X(\theta_1,\theta_2)=\bigl((R+r\cos\theta_1)\cos\theta_2,(R+r\cos\theta_1)\sin\theta_2,r\sin\theta_1\bigr),
%\end{equation}
%the orthogonal scale factors are $r$ and $R+r\cos\theta_1$. Standardizing these separately by $r$ and $R$ gives $1$ and $1+(r/R)\cos\theta_1$. We therefore use
%\begin{equation}
%\label{eq:t2metric}
%\dd\widetilde s_\lambda^2=\dd\theta_1^2+(1+\lambda\cos\theta_1)^2\dd\theta_2^2.
%\end{equation}
%For a physical curved torus $\lambda=r/R$; in the statistical family, $\lambda$ is a \emph{geometry} or \emph{warping} parameter. Because the coordinate scale factors are standardized individually, \eqref{eq:t2metric} should be viewed as a statistical construction inspired by the induced metric, not as the induced metric divided by a global radius scale.
%
%For the recursive construction of \citet{llibre2024}, let the standardized scale factors satisfy
%\begin{equation}
%\label{eq:hk}
%\dd\widetilde s_\lambda^2=\sum_{k=1}^N h_k(\theta;\lambda)^2\dd\theta_k^2,
%\quad h_1=1,\quad h_k=1+\lambda h_{k-1}\cos\theta_{k-1}.
%\end{equation}
%The multi-parameter coordinatewise standardization from the underlying recursive radii is derived in Appendix~\ref{app:metric-normalization}.

\begin{theorem}
\label{thm:recursive-positivity}
For $k\ge1$ define $b_k(\lambda)=\sum_{j=0}^{k-1}\lambda^j$ and $a_k(\lambda)=1-\sum_{j=1}^{k-1}\lambda^j$. If $\lambda_0\ge0$ satisfies
\begin{equation}
\label{eq:uniform-positivity-condition}
\sum_{j=1}^{N-1}\lambda_0^j<1,
\end{equation}
then for every $\lambda\in[0,\lambda_0]$, $\theta\in\T^N$, and $k\le N$,
\begin{equation}
\label{eq:recursive-uniform-bounds}
0<\underline h_{N,\lambda_0}\le a_k(\lambda)\le h_k(\theta;\lambda)\le b_k(\lambda)\le\overline h_{N,\lambda_0},
\end{equation}
where $\underline h_{N,\lambda_0}=1-\sum_{j=1}^{N-1}\lambda_0^j$ and $\overline h_{N,\lambda_0}=\sum_{j=0}^{N-1}\lambda_0^j$. Hence
\begin{equation}
\label{eq:uniform-ellipticity}
\underline h_{N,\lambda_0}^2\|v\|_2^2\le g_{\lambda,\theta}(v,v)\le \overline h_{N,\lambda_0}^2\|v\|_2^2.
\end{equation}
	Hence $\{g_\lambda:0\le\lambda\le\lambda_0\}$ is uniformly positive
definite on $\T^N$.
The condition is sufficient, not necessary.
\end{theorem}

The recursive geometry is intentionally order-dependent, $h_k$ depends on preceding coordinates. It is also tied to the chosen angular origins. A phase relabelling $\theta_j^\star=\theta_j+\alpha_j$ generally changes the fixed-form coefficients in \eqref{eq:hk}; invariance is recovered only when the observations and the metric are transformed together by the corresponding pullback. Thus coordinate order and angular reference origins are part of the geometric specification and must be fixed from scientific conventions rather than tuned using cluster labels. In the real-data analysis we retain the supplied torsion-angle conventions.

	\subsection{Energy over lifted coordinate-linear paths}
\label{sec:cte-energy}

Let $x=[\theta]=(\theta_1,\theta_2,\cdots, \theta_N)$ and \\ $y=[\eta]=(\eta_1,\eta_2,\cdots, \eta_N)$ be two points in $\T^N$, and define the
coordinatewise principal angular difference \\$	\delta
=
\wrap(\eta-\theta)
\in(-\pi,\pi]^N.$
For a winding vector $m\in\Z^N$, define $	\Delta^{(m)}
=
\delta+2\pi m.$
The corresponding coordinate-linear lift in $\R^N$ is
\[
\widetilde{\gamma}_m(t)
=
\theta+t\Delta^{(m)},
\quad
0\le t\le1,
\]
whose projection onto the torus is $	\gamma_m(t)
=
[\widetilde{\gamma}_m(t)]
\in\T^N.$
Thus $\gamma_m(0)=x$ and $\gamma_m(1)=y$, while different
$m\in\Z^N$ represent different winding classes connecting the same
pair of torus points.
Now, for an absolutely continuous path
$\gamma:[0,1]\to\T^N$, the standard Riemannian path-energy functional
associated with the metric $g_\lambda$ is
\begin{equation}
	\label{eq:riemannian-energy}
	E_{g_\lambda}[\gamma]
	=
	\int_0^1
	g_{\lambda,\gamma(t)}
	\bigl(\dot{\gamma}(t),\dot{\gamma}(t)\bigr)
	\,\dd t .
\end{equation}
This is the usual Riemannian energy functional; the corresponding path
length is obtained by integrating the square root of the same quadratic
form which is the standard Riemannian path-energy functional
\citep[Section~3.2]{srivastava2016functional}.  In particular, the energy
functional itself is not new here.  Our construction restricts the
standard Riemannian energy to the computationally tractable family of
coordinate-linear lifts indexed by winding classes.

For the path $\gamma_m$, the lifted velocity is constant,  $	\dot{\widetilde{\gamma}}_m(t)
=
\Delta^{(m)}.$
Hence, under the recursive metric \eqref{eq:hk}, its Riemannian energy
takes the explicit form
\begin{equation}
	\label{eq:pathenergy}
	E_m(x,y)
	=
	E_{g_\lambda}[\gamma_m]
	=
	\int_0^1
	\sum_{k=1}^N
	h_k\bigl(\widetilde{\gamma}_m(t);\lambda\bigr)^2
	\bigl(\Delta_k^{(m)}\bigr)^2
	\,\dd t .
\end{equation}
Here the metric coefficients are evaluated periodically, so the value is
independent of the chosen representatives of the torus coordinates.

We then define the curved-torus energy by minimizing over all winding
classes,
\begin{equation}
	\label{eq:cte}
	\E(x,y)
	=
	\min_{m\in\Z^N}E_m(x,y),
	\quad
	\DE(x,y)
	=
	\sqrt{\E(x,y)}.
\end{equation}
The minimization is therefore not over all admissible paths on the
Riemannian manifold, but only over coordinate-linear lifted paths and
their winding classes.  Consequently,\textit{ CTE is a computable
geometry-aware dissimilarity rather than the exact Riemannian geodesic
distance.}

%\subsection{Energy over lifted coordinate-linear paths}
%\label{sec:cte-energy}
%For $x=[\theta]$ and $y=[\eta]$, let $\delta=\wrap(\eta-\theta)\in(-\pi,\pi]^N$ and $\Delta^{(m)}=\delta+2\pi m$ for $m\in\Z^N$. The coordinate-linear lift is $\widetilde\gamma_m(t)=\theta+t\Delta^{(m)}$, and its Riemannian energy is
%\begin{equation}
%\label{eq:pathenergy}
%E_m(x,y)=\int_0^1\sum_{k=1}^N h_k(\widetilde\gamma_m(t);\lambda)^2\bigl(\Delta_k^{(m)}\bigr)^2\dd t.
%\end{equation}
%We define
%\begin{equation}
%\label{eq:cte}
%\E_\lambda(x,y)=\min_{m\in\Z^N}E_m(x,y),
%\quad \DEpar{\lambda}(x,y)=\sqrt{\E_\lambda(x,y)}.
%\end{equation}
%Thus CTE minimizes the standard Riemannian energy \citep{srivastava2016functional} over coordinate-linear lifted paths and winding classes, not over all manifold paths.

% \begin{theorem}
% \label{thm:coordinate_relabelling}
% Let $F_\pi([\theta])=[P_\pi\theta]$ for a coordinate permutation $\pi$ and let $g^{(\pi)}=(F_\pi^{-1})^*g$. Then
% $$
% \DE^{\,g^{(\pi)}}\!\bigl(F_\pi(x),F_\pi(y)\bigr)=\DE^{\,g}(x,y).
% $$
% \end{theorem}

\begin{theorem}
\label{thm:coordinate_relabelling}
Let $\pi$ be a coordinate permutation, $P_\pi$ its permutation matrix,
and
$
F_\pi([\theta])=[P_\pi\theta].
$
Let $g^{(\pi)}=(F_\pi^{-1})^*g$ be the relabeled metric, and let
$\DE^g$ denote the CTE dissimilarity under $g$. Then
$$
\DE^{\,g^{(\pi)}}\!\bigl(F_\pi(x),F_\pi(y)\bigr)
=
\DE^{\,g}(x,y).
$$
\end{theorem}

\begin{remark}
\label{rem:permutation}
Holding the recursive metric fixed while permuting only the data generally changes CTE; coordinate order is therefore part of the model.
\end{remark}

The definition \eqref{eq:cte} appears to require minimization over the
infinite lattice $\Z^N$. The next result shows that strict positivity of
the metric coefficients reduces this optimization to an explicit
finite set of winding classes.

\begin{theorem}
\label{thm:finite}
Suppose $h_k(\vartheta)\ge a_k >0$ uniformly and let $U=E_0(x,y)$. Every minimizing winding vector $m^\star$ satisfies
\begin{equation}
\label{eq:ellipsoid}
\sum_{k=1}^N a_k^2(\delta_k+2\pi m_k^\star)^2\le U,
\end{equation}
so each coordinate lies in the finite interval
\begin{equation}
\label{eq:windinterval}
\left\lceil\frac{-\sqrt U/a_k-\delta_k}{2\pi}\right\rceil
\le m_k^\star\le
\left\lfloor\frac{\sqrt U/a_k-\delta_k}{2\pi}\right\rfloor.
\end{equation}
Defining
\begin{equation}
\label{eq:mkbound}
M_k(U)=\left\lfloor\frac{\sqrt U/a_k+\pi}{2\pi}\right\rfloor,
\end{equation}
the number of candidates is bounded by
\begin{equation}
\label{eq:candidatebound}
|\mathcal M_U(x,y)|\le\prod_{k=1}^N(2M_k(U)+1).
\end{equation}
For the recursive family \eqref{eq:hk}, a minimizer can be chosen with $m_N=0$, so the canonical search satisfies
\begin{equation}
\label{eq:canonicalcount}
|\mathcal M_U^{\rm can}(x,y)|\le\prod_{k=1}^{N-1}(2M_k(U)+1).
\end{equation}
Hence exact winding minimization within the coordinate-linear path family reduces to a certified finite search.
\end{theorem}

\begin{corollary}
\label{cor:uniform-winding}
Under Theorem~\ref{thm:recursive-positivity}, the principal-path bound
\begin{equation}
\label{eq:uglobal}
U\le\overline U_{N,\lambda}=\pi^2\sum_{j=1}^N b_j^2 
\end{equation}
yields a pair-uniform finite candidate bound
\begin{equation}
\label{eq:cglobal}
\overline C_{N,\lambda}=\prod_{k=1}^{N-1}(2\overline M_k+1),
\quad
\overline M_k=\left\lfloor\frac{\sqrt{\overline U_{N,\lambda}}/a_k+\pi}{2\pi}\right\rfloor.
\end{equation}
For fixed $N$ and $\lambda$ in a compact subset of the positivity region, $\overline C_{N,\lambda}$ is uniformly bounded.
\end{corollary}

\begin{proposition}
\label{prop:cte-complexity}
With $Q$-node Gauss-Legendre quadrature, one CTE pair has worst-case arithmetic cost $O(QN\overline C_{N,\lambda})$, and a dense $n\times n$ CTE matrix costs $O(n^2QN\overline C_{N,\lambda})$. These are completeness bounds, not claims that $\overline C_{N,\lambda}$ stays small near the positivity boundary or as $N$ grows.
\end{proposition}

\begin{theorem}
\label{thm:cte-properties}
If $0<a\le h_k(\theta;\lambda)\le b<\infty$, then CTE is well defined, lift-invariant, symmetric, nonnegative, and zero only at identical torus points. With $d_0$ the flat product-torus distance and $d_{g_\lambda}$ the Riemannian geodesic distance,
\begin{equation}
\label{eq:cte-geometric-comparison}
a\,d_0(x,y)\le d_{g_\lambda}(x,y)\le\DEpar{\lambda}(x,y)\le b\,d_0(x,y).
\end{equation}
At $\lambda=0$,
\begin{eqnarray}
\label{eq:cte-flat-endpoint}
\DEpar{0}(x,y) &=& d_{g_0}(x,y) \nonumber\\
& = &d_0(x,y)
= \left\{\sum_{k=1}^N\wrap(\eta_k-\theta_k)^2\right\}^{1/2}
\end{eqnarray}
\end{theorem}

\begin{remark}
\label{rem:cte-not-metric}
CTE need not satisfy the triangle inequality because the energy minimization is restricted to coordinate-linear lifted paths; Appendix~\ref{app:triangle} gives an explicit counterexample.
\end{remark}

\begin{theorem}[Stability in the warping parameter]
\label{thm:lambda-stability}
On every compact $[0,\lambda_0]$ satisfying Theorem~\ref{thm:recursive-positivity}, finite constants $C_{N,\lambda_0}$ and $L_{N,\lambda_0}$ exist such that
\begin{equation}
\label{eq:energy-multiplicative-stability}
e^{-C_{N,\lambda_0}|\lambda-\mu|}\E_\lambda(x,y)
\le\E_\mu(x,y)\le
e^{C_{N,\lambda_0}|\lambda-\mu|}\E_\lambda(x,y),
\end{equation}
and
\begin{equation}
\label{eq:distance-lipschitz}
|\DEpar{\mu}(x,y)-\DEpar{\lambda}(x,y)|\le L_{N,\lambda_0}|\mu-\lambda|.
\end{equation}
%Explicit constants and proofs are in Appendix~\ref{app:lambda-stability-proof}.
\end{theorem}

\section{Clustering and predictive geometry selection} 
\label{sec:cehs}

For fixed $K$, let $m_1,\ldots,m_K$ denote the cluster medoids and
let $z_i\in\{1,\ldots,K\}$ denote the cluster assignment of
observation $x_i$. Then energy $K$-medoids minimizes
$$
\sum_{i=1}^n \DE(x_i,m_{z_i})^2
=
\sum_{i=1}^n \E_\lambda(x_i,m_{z_i}).
$$
For a candidate $\lambda\in\Lambda$, energy $K$-medoids is fitted
on a training split, producing training medoids
$m_1,\ldots,m_K$. Each validation observation is then assigned to
its nearest training medoid under $\DE$. Let $\tau_\lambda$ denote
the median positive pairwise energy among the training observations,
and define
$$
q_\lambda(x,y)
=
\exp\left\{
-\frac{\E_\lambda(x,y)}{\tau_\lambda}
\right\}.
$$
If $\widehat H_c$ is the validation-cluster mean similarity to its training medoid, let $\widehat H_{.25}=Q_{.25}(\widehat H_1,\ldots,\widehat H_K)$.
For medoid pairs define
$R_{cd}=1-q_\lambda(m_c,m_d)$ and
$$
Q_{\rm HM}
=
\frac{M}{\sum_{c<d}R_{cd}^{-1}},
\quad
M=K(K-1)/2,
$$
with $Q_{\rm HM}=0$ if any medoid pair is unseparated.
A held-out evaluation is declared invalid if any fitted training
cluster receives fewer than two validation observations.  Such an
evaluation is assigned CEHS-HM score zero.  Thus, for held-out
evaluation $\ell$, we use $\displaystyle \widehat C_{\lambda,K,\ell}
=
	\dfrac{2\widehat H_{.25,\ell}Q_{{\rm HM},\ell}}
	{\widehat H_{.25,\ell}+Q_{{\rm HM},\ell}},
$ 	if every fitted cluster receives at least two validation
	observations, and  $\widehat C_{\lambda,K,\ell}
	=0$ otherwise.
This zero-score convention penalizes candidate geometries and
resolutions that produce empty or nearly empty held-out clusters,
rather than removing their unfavorable evaluations from the model
selection criterion.  
% The primary fixed $K$ selector is

% \begin{equation}
% \label{eq:lambda-selector}
% \widehat\lambda(K)\in\arg\max_{\lambda\in\Lambda}\widehat C_{\lambda,K}.
% \end{equation}
% The selected $\widehat\lambda(K)$ is a clustering-geometry tuning parameter; it is not defined as an estimator of a physical or generative deformation parameter. Joint maximization over $(\lambda,K)$ is retained only as a secondary extension in Appendix~\ref{app:cehs-joint}.  \vspace{-.5em}

\paragraph{Geometry-selection workflow.}
For a specified clustering resolution $K$, CEHS geometry selection proceeds as follows.
For each $\lambda\in\Lambda$ and each prespecified train-validation split:
(i) fit energy $K$-medoids to the training observations using $E_\lambda$;
(ii) compute the training scale $\tau_\lambda$ as the median positive
pairwise training energy;
(iii) assign validation observations to their nearest fitted medoids;
(iv) compute $\widehat H_{.25,\ell}$,
$Q_{{\rm HM},\ell}$, and
$\widehat C_{\lambda,K,\ell}$ from the held-out observations; and
(v) average the CEHS-HM scores over all prespecified held-out
evaluations, including the zero scores assigned to invalid
evaluations.
If $L$ held-out evaluations are prespecified, the resulting criterion is
$$
\widehat C_{\lambda,K}
=
\frac{1}{L}
\sum_{\ell=1}^{L}
\widehat C_{\lambda,K,\ell}.
$$

The selected geometry is
$
\widehat\lambda(K)
\in
\arg\max_{\lambda\in\Lambda}
\widehat C_{\lambda,K}.
$
The same held-out splits are reused for every candidate geometry, so
the comparison across $\lambda$ is paired.

When the clustering resolution $K$ is not specified externally, the
same zero-penalized held-out criterion can be extended to a
prespecified finite joint grid,
$$
(\widehat\lambda,\widehat K)
\in
\arg\max_{(\lambda,K)\in\mathcal G}
\widehat C_{\lambda,K}.
$$
We use this only as a secondary data-adaptive extension, for example,
in real-data applications such as ILE, where no clustering resolution
is externally specified. The resulting $\widehat K$ and
$\widehat\lambda$ are not interpreted as estimators of generative
parameters or of a generative number of mixture components.

\begin{proposition}
\label{prop:cehs}
For a valid split, CEHS-HM lies in $[0,1]$ and is invariant to cluster-label permutations and to multiplication of all energies at fixed $\lambda$ by a common positive constant. Conditional on the training fit, the score is strictly increasing in $\widehat H_{.25}$ when $Q_{\rm HM}>0$ is fixed and strictly increasing in $Q_{\rm HM}$ when $\widehat H_{.25}>0$ is fixed. For positive separations $R_1,\ldots,R_M$,\\
$\displaystyle
\min_jR_j\le Q_{\rm HM}\le M^{-1}\sum_{j=1}^MR_j.
$
No triangle inequality is required.
\end{proposition}
These are properties of the validation criterion; they do not imply consistency for a generative $\lambda_0$ or a generative mixture-component count.
For the concentration analysis, fix $K$ and suppress it from the
notation.  For held-out evaluation $\ell$, write
$\widehat C_{\lambda\ell}
\equiv
\widehat C_{\lambda,K,\ell}$.
Conditioning on the corresponding training fit, let
$C_{\lambda\ell}$ denote the population counterpart obtained by
replacing the empirical validation homogeneities by their conditional
expectations.  Define
$
\widehat C_\lambda
=
\frac{1}{L}\sum_{\ell=1}^L
\widehat C_{\lambda\ell},
\quad
C_\lambda
=
\frac{1}{L}\sum_{\ell=1}^L
C_{\lambda\ell}.
$

These properties characterize the behavior of the held-out
criterion but do not imply recovery of a generative deformation
parameter or mixture-component count. We therefore assess the
finite-sample behavior of the CEHS selector directly through the
controlled and neutral simulation studies in Section~\ref{sec:experiments}.

% \begin{theorem}
% \label{thm:cehs-concentration}
% Fix $K\ge2$ and a finite $\Lambda$. If observations are i.i.d. and validation observations are disjoint from the corresponding training sample, define
% $$
% \mathcal V_r=\left\{\min_{\lambda\in\Lambda}\min_{1\le\ell\le L}\min_{1\le c\le K}n_{\lambda\ell,c}\ge r\right\},\quad r\ge2.
% $$
% Then, without requiring independence across repeated held-out evaluations,
% \begin{equation}
% \label{eq:cehs-uniform-concentration}
% \Pr\!\left(\mathcal V_r\cap\left\{\max_{\lambda\in\Lambda}|\widehat C_\lambda-C_\lambda|>t\right\}\right)
% \le2LK|\Lambda|e^{-rt^2/2}.
% \end{equation}
% If $\widehat\lambda\in\arg\max_\lambda\widehat C_\lambda$ and $\lambda^\star\in\arg\max_\lambda C_\lambda$, then for every $\delta\in(0,1)$,
% \begin{equation}
% \label{eq:cehs-oracle}
% \Pr\!\left[\mathcal V_r\cap\left\{C_{\lambda^\star}-C_{\widehat\lambda}>2\sqrt{\frac{2}{r}\log\!\left(\frac{2LK|\Lambda|}{\delta}\right)}\right\}\right]\le\delta.
% \end{equation}
% \end{theorem}

% The event $\mathcal V_r$ is used only for the concentration statement.
% Outside this event, the implemented model-selection criterion remains
% well defined through the zero-score convention above; no invalid
% held-out evaluation is discarded when computing the empirical
% selection criterion. 

% Here $\lambda^\star$ maximizes the population held-out CEHS criterion on the candidate grid; it need not equal a parameter used to generate the observations. Theorem~\ref{thm:cehs-concentration} is therefore a geometry-selection regret guarantee, not a generative-parameter consistency theorem.

\section{Experiments}  
\label{sec:experiments}

\subsection{Clustering artificial toroidal data}
\label{sec:artificial_clustering}

We assess comparative clustering performance using controlled artificial
toroidal data for which the generating cluster memberships are known.
This separation between simulation-based method comparison and
real-data illustration is standard in unsupervised clustering, since
external performance measures such as the adjusted Rand index require a
reference partition that is generally unavailable for genuinely
unlabelled real data.

A similar evaluation strategy is adopted by
\citet{jung2021clustering}: their real protein torsion-angle data are
used to illustrate and interpret the recovered cluster structure,
whereas empirical comparisons with competing clustering procedures are
carried out on artificial toroidal data with known cluster membership.
We follow the same principle here.  The simulation experiments are used
for controlled comparisons of clustering accuracy, geometry selection,
and robustness, while the real-data analyses are used to assess
stability and scientifically interpretable structure without treating
external annotations as ground-truth cluster labels.

\subsubsection{\texorpdfstring{$\T^2$}{T2} morphology benchmark}  
We consider two complementary classes of artificial toroidal data.
The first consists of compact curved basins, designed to favour
localized cluster structure.  The
second consists of nonlinear curved bands, which provide a more
challenging nonconvex morphology.  Both designs contain $K=3$ clusters
with $n=300$ observations and are generated under curvature
$\lambda=0.75$. The same CTE geometry is supplied to several engines because compact basins and elongated bands need not favor the same clustering objective. Table~\ref{tab:t2} shows the resulting morphology dependence, center/medoid methods are strongest for basins, whereas HDBSCAN is strongest for bands. PSM-DBSCAN and MINT-AGE are secondary reimplementations, MINT-AGE also estimates its own cluster number. The full method table, paired PSM-DBSCAN contrasts, benchmark figure, and official ClusTorus comparison are reported in Appendix~\ref{app:t2_details}.

For the basin design, the Fr\'echet-center and energy-medoid
procedures have the largest mean ARIs, $0.769$ and $0.767$,
respectively. For the band design, CTE-HDBSCAN has the largest
mean ARI, $0.578$, although its paired contrast with PSM-DBSCAN
is not separated from zero. Additional $\T^2$ morphology benchmarks and complete-procedure comparisons with ClusTorus, PSM-DBSCAN, and a dimension-general MINT-AGE reimplementation are
reported in Appendix~\ref{app:t2_details}.

	\begin{table}[h!]
	\centering
	\caption{$\mathbb{T}^2$ mean ARI (SD) over $30$ paired replicates.}
	\label{tab:t2}
	\scalebox{1.2}{\begin{tabular}{lcc}
		\toprule
		Method & Curved basins & Curved bands \\
		\midrule
		CTE Fr\'echet K-means & 0.769 (0.055) & 0.374 (0.086) \\
		CTE energy-medoid      & 0.767 (0.053) & 0.349 (0.085) \\
		CTE spectral           & 0.714 (0.072) & 0.335 (0.131) \\
		CTE hierarchical       & 0.672 (0.149) & 0.419 (0.137) \\
		CTE HDBSCAN            & 0.605 (0.196) & 0.578 (0.203) \\
		CTE DBSCAN             & 0.256 (0.280) & 0.494 (0.074) \\
		PSM-DBSCAN             & 0.544 (0.196) & 0.520 (0.115) \\
		Flat DBSCAN            & 0.229 (0.294) & 0.485 (0.061) \\
		MINT-AGE reimpl.       & 0.414 (0.209) & 0.203 (0.074) \\
		\bottomrule
	\end{tabular}}
 
\end{table}

\subsection{Controlled geometry ablation}  
We next isolate the geometry by holding the clustering engine fixed. Data are generated under \eqref{eq:hk} in $\T^3$ at $\lambda\in\{0,.15,.30,.45,.55\}$, with $K=3$, $n=300$, and 100 paired replicates per level. CTE and flat energy-medoid clustering use the same initialization schedule, and the same latent seed is reused $\lambda$ within a replicate.

\begin{table}[h]
	\centering
	\caption{Paired curvature ablation.\\ $\Delta=\ARI_{\rm CTE}-\ARI_{\rm flat}$ over 100 replicates.}
	\label{tab:curvature}
	\scalebox{1.2}{	\begin{tabular}{c@{\hspace{5pt}}cc@{\hspace{5pt}}c}
		\toprule
		$\lambda$ & CTE ARI & Flat ARI & $\Delta$ (95\% CI)\\
		\midrule
		0.00 & .8199 & .8199 & $.0000\ [0,0]$\\
		0.15 & .8690 & .8687 & $.0002\ [-.0026,.0031]$\\
		0.30 & .9149 & .9068 & $.0081\ [.0042,.0121]$\\
		0.45 & .9454 & .9266 & $.0189\ [.0146,.0231]$\\
		0.55 & .9594 & .9364 & $.0230\ [.0189,.0271]$\\
		\bottomrule
	\end{tabular}}
\end{table}

\begin{figure}[t]
	\centering
	\includegraphics[width=0.9\columnwidth]{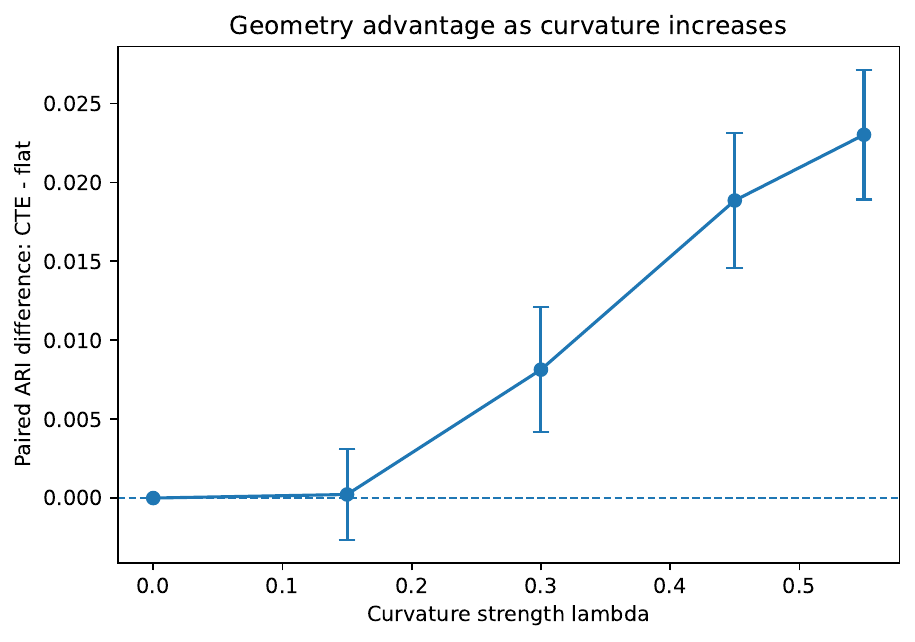}
	\caption{Paired CTE-minus-flat ARI versus warping. Error bars are 95\% Monte Carlo intervals.}
	\label{fig:curvature}
    
\end{figure}

At $\lambda=0$, Theorem~\ref{thm:cte-properties} forces the procedures coincide, and the empirical paired difference is exactly zero. Mild warping ($\lambda=.15$) is indistinguishable from zero; from $\lambda=.30$ onward the paired interval is positive and the advantage grows with warping (Table~\ref{tab:curvature}, Figure~\ref{fig:curvature}). Because only the geometry changes, this is the primary mechanism experiment.

\subsection{What CEHS selects: controlled and neutral studies}

In both studies, every candidate geometry is evaluated using the
CTE energy $K$-medoids procedure of Section~\ref{sec:cehs}; thus,
the selection experiments vary the candidate geometry and, only in
the secondary joint analysis, the clustering resolution $K$, while
keeping the clustering engine fixed.

The first final validation study examines the held-out selector itself
on 500 controlled three-cluster datasets, 100 replicates for each
generating deformation
$\lambda_0\in\{0,.15,.35,.45,.525\}$.
The correct resolution $K=3$ was selected in all 500 datasets, but
exact matching of $\widehat\lambda$ to the generating $\lambda_0$
was only $6\%,3\%,2\%,3\%$, and $82\%$, respectively; mean
absolute errors were $.424,.296,.151,.080$, and $.018$. In
particular, when $\lambda_0=0$, $94\%$ of joint selections used a
positive $\lambda$. The mean CEHS profile at $K=3$ also increased
modestly toward the upper candidate range even in this flat-generating
case. This confirms that $\widehat\lambda$ should be interpreted as
a clustering-geometry tuning parameter, not as an estimator of the
generating deformation. Full confusion matrices and score surfaces
are in Appendix~\ref{app:new-controlled-selection}.

The second study uses an equal-weight, three-component product-von-Mises mixture on $\T^3$, with independent coordinates within components, $\kappa=2$, $n=300$, and a common random toroidal phase shift in each of 100 replicates. No CTE geometry is used in sampling. With $K=3$ fixed, the selector chose $\widehat\lambda=0$ in $56\%$ of replicates and $\widehat\lambda\le.15$ in $74\%$; the median selected value was zero. The adaptive mean ARI was $0.8072$ versus $0.8138$ at the exact flat endpoint, giving paired difference $-0.0066$ (95\% CI $[-0.0118,-0.0014]$); $84\%$ of replicates were within $0.02$ ARI of the flat method. Thus the selector largely moves toward the flat end under this misspecification, with a small but nonzero average loss.

\begin{table}[t]
\centering
\caption{Neutral product-von-Mises $\T^3$ study with $K=3$ fixed (100 replicates).}
\label{tab:neutral}
\small
\scalebox{1.18}{\begin{tabular}{lc}
\toprule
Quantity & Result \\
\midrule
$P(\widehat\lambda=0)$ & 0.56 \\
$P(\widehat\lambda\le .15)$ & 0.74 \\
Mean ARI, flat endpoint & 0.8138 \\
Mean ARI, CEHS-selected geometry & 0.8072 \\
Paired ARI difference & $-0.0066$ \\
95\% CI for difference & $[-0.0118,-0.0014]$ \\
\bottomrule
\end{tabular}} 
\end{table}

Joint $(\lambda,K)$ optimization is not used as the primary claim.
On the same neutral data-generating process (DGP), it selected the
generating $K=3$ in only $6\%$ of replicates (mean selected
$K=2.06$) and had a paired ARI difference of $-0.3658$ relative
to flat $K=3$. This behavior is detailed in
Appendix~\ref{app:neutral-study} and motivates treating joint
resolution selection as a secondary finite-grid, data-adaptive
extension rather than as an estimator of a generative
mixture-component count. The earlier $1{,}000$-replicate
confirmatory comparison and the $\T^2$ morphology benchmark are
reported in Appendices~\ref{app:baseline-criteria} and
\ref{app:t2_details}, respectively; they are secondary to the
geometry-focused claims.

\subsection{Comparison with held-out baseline criteria}
\label{sec:criterion-comparison} 

We compare CEHS-HM with a held-out silhouette
\citep{rousseeuw1987silhouettes} and a pairwise
energy-dispersion pseudo-$F$ criterion motivated by
between-to-within dispersion ratios
\citep{calinski1974dendrite,anderson2001new}.
All three criteria are evaluated on the same candidate grid,
training fits, held-out splits, and validation assignments.
Each criterion independently selects its own
$(\widehat\lambda,\widehat K)$ by maximizing its mean held-out
score. Full definitions and implementation details are given in
Appendix~\ref{app:baseline-criteria}.

\begin{table}[h]
	\centering
	\caption{Aggregate confirmatory comparison of three label-free
		model-selection criteria over $1000$ fresh replicates.
		ARI regret is the independent-test oracle ARI minus the ARI of
		the selected candidate; severe regret denotes regret at least
		$0.10$. Lower regret is better.}
	\label{tab:cehs_confirmatory_main}
	\small
	\scalebox{1.186}{\begin{tabular}{lccc}
		\toprule
		Criterion &
		$P(\widehat K=K_0)$ &
		Mean ARI regret &
		Severe regret \\
		\midrule
		CEHS-HM   & $0.988$ & $0.0263$ & $0.037$ \\
		Pseudo-$F$ & $0.947$ & $0.0325$ & $0.077$ \\
		Silhouette & $0.922$ & $0.0378$ & $0.093$ \\
		\bottomrule
	\end{tabular}}
       
\end{table}

The  confirmatory experiment contains $250$ fresh replicates
in each of four prespecified scenarios, giving $1000$ replicates in
total.  
CEHS-HM selects the generating cluster resolution in $98.8\%$ of
replicates and has mean independent-test ARI regret $0.0263$.
The corresponding values are $94.7\%$ and $0.0325$ for pseudo-$F$,
and $92.2\%$ and $0.0378$ for silhouette. The prespecified paired
comparison with pseudo-$F$ gives a CEHS-HM minus pseudo-$F$
mean-regret difference of $-0.00622$, with a $95\%$ stratified
paired-bootstrap interval $[-0.00965,-0.00276]$. Thus the aggregate comparison in
Table~\ref{tab:cehs_confirmatory_main} should not be interpreted as
uniform superiority of CEHS-HM across every data-generating
configuration
The scenario-level results in Appendix~\ref{app:baseline-criteria}
show that the aggregate advantage is not uniform across all
settings.

\subsection{Geodesic robustness and computational scaling} \vspace{-.5em}

Because CTE restricts the path class, we first compare it directly with
numerical Riemannian geodesic optimization. Across 2,000 uniformly sampled
full-torus pairs, Spearman correlations between CTE and optimized geodesic
distance are $0.9950$ on $\T^2$ and $0.9890$ on $\T^3$, with median
relative gaps $5.48\%$ and $4.49\%$. On 500 deliberately difficult winding
pairs, the corresponding correlations remain $0.9905$ and $0.9663$, with
median gaps $8.15\%$ and $6.19\%$. Across all 5,000 global and difficult
pairs, there are no numerical violations of $d_g\le D_E$. Thus CTE acts as
a computable upper-bound surrogate that closely preserves geodesic ordering,
rather than as an exact geodesic distance. 

We next examine empirical scaling with the torus dimension. To complement
the worst-case bound in Proposition~\ref{prop:cte-complexity}, we consider
$N\in\{2,3,5,7,10\}$ with $n=300$, $K=3$, and $\lambda=.40$. The mean
time required to construct a dense CTE distance matrix increases from
$0.849$ s at $N=2$ to $2.204$ s at $N=10$. In these compact-cluster
experiments, the maximum observed number of winding candidates per pair is
$2,2,5,7,$ and $13$, respectively. These measurements are empirical rather
than worst-case guarantees; candidate growth can be substantially larger
near the positivity boundary or for more difficult winding configurations.
Table~\ref{tab:scaling-main} summarizes the dimension-wise results.
Full diagnostics are given in
Appendix~\ref{app:geodesic}.
\begin{table}[t]
	\centering
	\caption{Empirical dimension scaling ($n=300$, 100 replicates).
		The geometry-effect column is paired CTE minus flat ARI.}
	\label{tab:scaling-main}
	\small
	\scalebox{1.18}{\begin{tabular}{rrrr}
		\toprule
		$N$ & $\Delta$ARI (95\% CI) & CTE matrix (s) &
		max.\ candidates \\
		\midrule
		2  & .0188 [.0128,.0248]   & .849 & 2 \\
		3  & .0311 [.0238,.0383]   & .950 & 2 \\
		5  & .0166 [.0088,.0244]   & 1.307 & 5 \\
		7  & .0067 [.0013,.0120]   & 1.646 & 7 \\
		10 & -.0052 [-.0106,.0001] & 2.204 & 13 \\
		\bottomrule
	\end{tabular}}
   
\end{table}

	\section{Real-data details} 
\subsection{ILE protein torsion data}
\label{sec:ile_application} 

We use the ILE data distributed with the \texttt{ClusTorus}
package \cite[see][]{hongjung2022} as the principal real-data application. The data contain
$n=8080$ complete observations, each represented by the four ordered
torsion angles
$(\phi,\psi,\chi_1,\chi_2)\in\mathbb{T}^4$.
The backbone angles $(\phi,\psi)$ and side-chain angles
$(\chi_1,\chi_2)$ are analyzed jointly. The coarse side-chain
rotamer-like sectors are not used in geometry selection or clustering;
they are retained only for post-hoc structural interpretation. Now,
for the geometry selection, the independently validated curved-geometry
grid is\\
$ \displaystyle
\Lambda_{\mathrm{curved}}
=
\{0.15,0.25,0.35,0.40,0.45,0.475,0.50,0.525\}.
$
For the ILE analysis we additionally include $\lambda=0$ as the exact
flat-product-torus reference established by
Theorem~\ref{thm:cte-properties}, giving
$$
\Lambda
=
\{0,0.15,0.25,0.35,0.40,0.45,0.475,0.50,0.525\}.
$$
The flat endpoint is therefore included as a reference candidate and
not presented as part of the independently validated curved-geometry
grid. Its inclusion does not alter the selected ILE model.

Because no clustering resolution $K$ is externally specified for the
ILE data, we use a secondary finite-grid extension of the CEHS-HM
procedure and select geometry and clustering resolution jointly,
$$
(\widehat\lambda,\widehat K)
\in
\arg\max_{\lambda\in\Lambda,\;K\in\{2,\ldots,20\}}
\widehat C_{\lambda,K}.
$$
This joint selection is used only as a data-adaptive choice of
clustering resolution and is not interpreted as estimation of a
generative number of mixture components. The same zero-penalized
held-out rule as in Section~\ref{sec:cehs} is used: if any fitted
cluster receives fewer than two validation observations in a
held-out evaluation, that evaluation is assigned CEHS-HM score zero.
All prespecified held-out evaluations, including such zero scores, are
retained when computing the mean criterion.

The selected pair is
$(\widehat\lambda,\widehat K)=(0.525,5)$.
The corresponding mean held-out CEHS-HM score is $0.81545$
(empirical repeated-split SE $0.00370$; 30 held-out evaluations).
For the selected pair, we generate ten CLARA-style clustering
solutions. Each solution is obtained by fitting energy $K$-medoids
on a $1000$-observation subsample and assigning all observations
to the resulting medoids. The mean pairwise ARI among the ten
resulting full-data partitions is $0.923$, with range
$0.855$-$0.9998$, indicating substantial subsampling stability
of the five-mode representation.

%For the final full-data partition at this selected pair, we use CTE
%energy $K$-medoids. Because $n=8080$, the full-data optimization is
%implemented through a CLARA-style approximation: ten candidate
%subsample fits are constructed and the candidate having the smallest
%full-data CTE energy objective is retained. Across the ten candidate
%fits, the mean pairwise adjusted Rand index is $0.923$, with range
%$0.855$-$0.9998$, indicating that the resulting five-mode
%representation is stable under repeated subsampling.

\begin{figure}[h!]
	\centering
	\includegraphics[width=\linewidth]{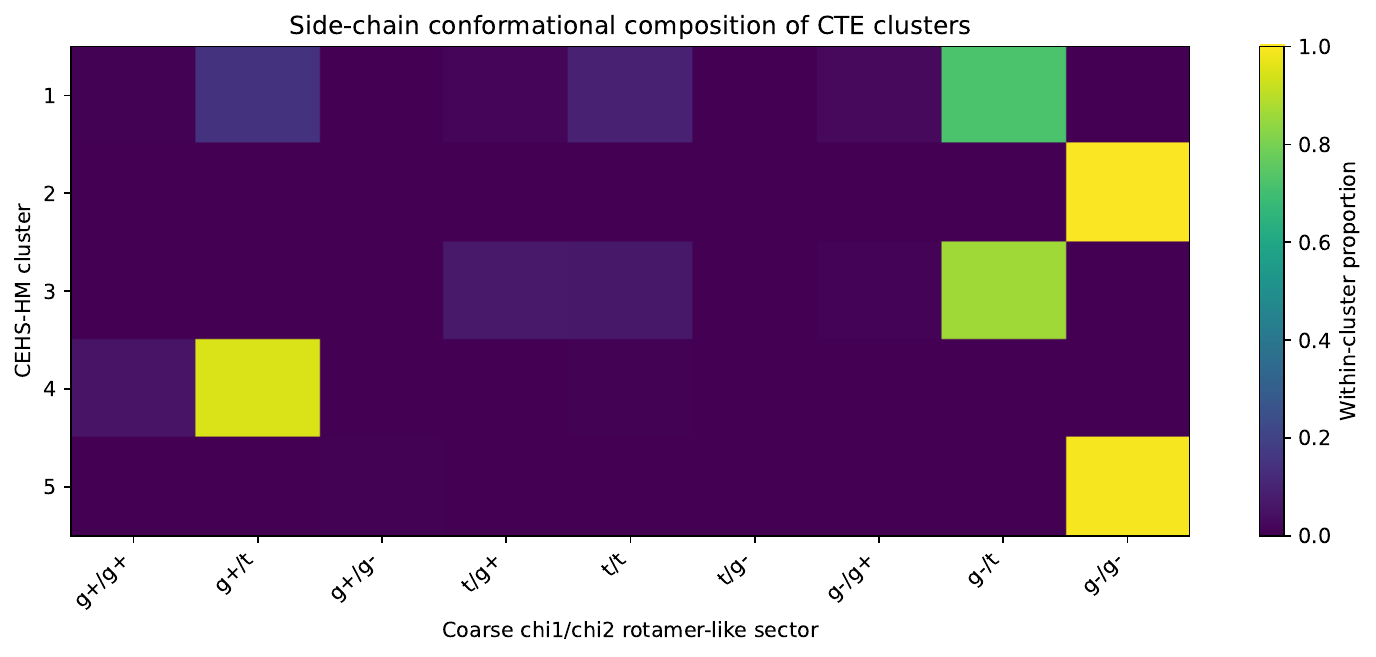}
	\caption{Composition of the five ILE conformational modes over the
			nine coarse $(\chi_1,\chi_2)$ side-chain sectors. These sectors are
			used only for post-hoc interpretation and are not supplied to the
			clustering procedure.}
	\label{fig:ile_composition}
\end{figure}

%The fitted clusters are interpreted as joint backbone-side-chain
%conformational modes rather than as externally labelled classes.
%This distinction is important because the same coarse side-chain
%sector can occur in different backbone environments.  For example,
%among the $4756$ observations in the $g^{-}/t$ sector, $2757$ belong
%to an extended-backbone mode and $1998$ to an $\alpha$-like mode.
%Likewise, the $1405$ observations in the $g^{-}/g^{-}$ sector split
%mainly between two different backbone regimes.  Thus the
%four-dimensional analysis identifies structure that is not captured
%by side-chain categorization alone.

Figure~\ref{fig:ile_composition} shows the within-cluster composition
of the five CEHS-HM conformational modes over the nine coarse
$(\chi_1,\chi_2)$ rotamer-like sectors.  In particular, it shows that
Clusters~1 and~3 are dominated by the $g^{-}/t$ sector, whereas
Clusters~2 and~5 are dominated by the $g^{-}/g^{-}$ sector, despite
their distinct backbone organization.  The corresponding contingency
counts, additional cluster summaries, and a post-hoc comparison with
the partition obtained by rerunning the published ClusTorus ILE
workflow are given in Appendix~\ref{app:ile_details}.
A second application to four-angle robotic-arm configurations is
reported in Appendix~\ref{app:omx}, providing a nonbiological example
in which the selected geometry also has nonzero curvature.

%The fitted clusters are interpreted as joint backbone-side-chain
%conformational modes rather than as externally labelled classes.
%This distinction is important because the same coarse side-chain
%sector can occur in different backbone environments.  For example,
%among the $4756$ observations in the $g^{-}/t$ sector, $2757$ belong
%to an extended-backbone mode and $1998$ to an $\alpha$-like mode.
%Likewise, the $1405$ observations in the $g^{-}/g^{-}$ sector split
%mainly between two different backbone regimes.  Thus the
%four-dimensional analysis identifies structure that is not captured
%by side-chain categorization alone.
%
%Figure~\ref{fig:ile_composition} summarizes the side-chain composition
%of the five modes.  Additional cluster summaries and a post-hoc comparison with the
%partition obtained by rerunning the published ClusTorus ILE workflow
%are given in Appendix~\ref{app:ile_details}.

%
%
%A second application to four-angle robotic-arm configurations is
%reported in Appendix~\ref{app:omx}, providing a nonbiological example
%in which the selected geometry also has nonzero curvature.

%\paragraph{Role of the real-data analysis.}
Unlike the controlled simulations, the ILE data do not provide a
defensible ground-truth cluster partition.  We therefore do not use
external agreement indices to rank competing methods on these data.
Instead, the real-data analysis is used to examine the stability and
structural interpretability of the partition selected by the proposed
label-blind procedure.  This distinction parallels the simulation and
real-data evaluation strategy used for toroidal clustering by
\citet{jung2021clustering}.

% \section{Discussion and conclusion} \vspace{-.8em}
% CTE addresses toroidal clustering when similarity may be position-dependent. Its main theoretical advantage is computability, the universal-cover formulation gives a certified finite winding search while preserving an exact flat endpoint and a direct Riemannian comparison. Its limitations are explicit, CTE is a path-restricted energy dissimilarity rather than an exact geodesic metric, and the recursive geometry depends on coordinate order and reference phases.

% The final experiments sharpen the statistical interpretation of
% CEHS-HM. The held-out selector is used for data-adaptive geometry
% selection, but its maximizer need not recover a parameter used to
% generate the observations. Likewise, joint finite-grid optimization
% over $(\lambda,K)$ may favor a coarser partition under overlapping
% mixtures. We therefore treat fixed-$K$ geometry selection as the
% primary formulation and use joint $(\lambda,K)$ selection only as a
% secondary practical extension when the clustering resolution is not
% specified externally, as in the ILE application.

% Empirically, the fixed-engine experiment shows that CTE offers increasing benefit as geometric warping strengthens, whereas the neutral product-von-Mises study shows concentration toward the flat end and only a small loss when curved geometry is unnecessary. Together with the ILE analysis, these results support CTE as a structured, interpretable way to separate toroidal geometry from clustering morphology while retaining the flat product torus as an exact special case.

% \newpage
\section{Discussion and conclusion}

We introduced curved-torus energy (CTE), a geometry-aware
dissimilarity for multivariate angular data on $\T^N$. The method uses
a recursive Riemannian metric with the flat product torus as the exact
endpoint $\lambda=0$. By working on the universal cover, CTE reduces
the path search to a certified finite set of winding vectors. We
established positivity, finite-winding reduction, computational
complexity, comparison with the Riemannian geodesic distance, and
stability in $\lambda$. CTE was then combined with energy $K$-medoids
and the held-out CEHS-HM criterion, and evaluated through simulation,
geodesic diagnostics, dimension scaling, and two real-data
applications.

CTE allows position-dependent toroidal geometry while remaining
computationally tractable. The finite winding search avoids unrestricted
geodesic optimization, and the exact flat endpoint provides a natural
reference when curvature is unnecessary. The experiments show improved
clustering under geometric warping, movement toward the flat geometry
under neutral misspecification, and strong preservation of geodesic
ordering. CEHS-HM also enables geometry selection without external
class labels.

CTE is a path-restricted energy dissimilarity rather than an exact
geodesic metric, and a triangle inequality is not guaranteed. The
recursive geometry depends on coordinate ordering and reference phases,
which should therefore be specified from scientific considerations.
The CEHS-HM maximizer is a data-adaptive geometry choice, not an
estimator of a generative deformation parameter, and joint
$(\lambda,K)$ optimization may favor coarser partitions. Computational
cost may also increase in higher dimensions or near the positivity
boundary.

Natural extensions include coordinate-specific deformation parameters,
structured dependency graphs, and principled treatment of coordinate
order and reference phases. Sharper winding bounds, pruning, and
parallel computation could improve scalability. Further work is also
needed on uncertainty quantification and statistical theory for
data-adaptive geometry selection, as well as applications to broader
classes of jointly periodic data.

% Overall, CTE provides a structured way to separate toroidal geometry
% from clustering morphology while retaining the flat product torus as
% an exact special case.
%\newpage
%\section*{AI Use Statement}
%Generative AI tools were used for literature discovery and
%summarization; assistance with the
%proof checking; design of simulation and real-data validation
%protocols; implementation, debugging, and refactoring of analysis
%code; interpretation of simulation and diagnostic outputs;
%preparation of reproducibility materials; and language editing and
%manuscript restructuring. All AI-assisted mathematical claims and
%proofs were checked by the authors, all reported computations were
%independently executed and verified against saved output files, and
%all cited sources were manually checked. The authors take full
%responsibility for the final content of the paper. Generative AI was
%not used to fabricate empirical observations or numerical results.

\bibliographystyle{apalike}
\bibliography{references}

@book{mardiajupp2000,
  author    = {Mardia, Kanti V. and Jupp, Peter E.},
  title     = {Directional Statistics},
  publisher = {Wiley},
  address   = {Chichester},
  year      = {2000},
  doi       = {10.1002/9780470316979}
}

@incollection{mescher2024geometric,
  title={Geometric and topological properties of manifolds in robot motion planning},
  author={Mescher, Stephan},
  booktitle={Topology and AI},
  pages={123--159},
  year={2024},
  publisher={European Mathematical Society-EMS-Publishing House GmbH}
}

@article{dou2026neuralactuator,
  title={NeuralActuator: Neural Actuation Modeling for Robot Dynamics and External Force Perception},
  author={Dou, Zhiyang and Onyemelukwe, John U and Zhang, Hangxing and Zhang, Heng and Guo, Minghao and Tian, Yunsheng and Lipiec, Michal Piotr and Jacob, Joshua and Liu, Chao and Chen, Peter Yichen and others},
  journal={arXiv preprint arXiv:2607.11734},
  year={2026}
}

@article{jha2017multiple,
  title={Multiple circular--circular regression},
  author={Jha, Jayant and Biswas, Atanu},
  journal={Statistical Modelling},
  volume={17},
  number={3},
  pages={142--171},
  year={2017},
  publisher={SAGE Publications Sage India: New Delhi, India}
}

@article{jha2018circular,
  title={Circular-circular regression model with a spike at zero},
  author={Jha, Jayant and Biswas, Atanu},
  journal={Statistics in Medicine},
  volume={37},
  number={1},
  pages={71--81},
  year={2018},
  publisher={Wiley Online Library}
}

@article{biswas2026efficient,
  title={Efficient sampling from circular distributions: extensions to toroidal and spherical distributions},
  author={Biswas, Surojit and Banerjee, Buddhananda},
  journal={Communications in Statistics-Simulation and Computation},
  pages={1--24},
  year={2026},
  publisher={Taylor \& Francis}
}

@article{biswas2024intrinsic,
  title={Intrinsic Geometry-Based Angular Covariance: A Novel Framework for Nonparametric Changepoint Detection in Meteorological Data},
  author={Biswas, Surojit and Banerjee, Buddhananda and Laha, Arnab Kumar},
  journal={arXiv preprint arXiv:2409.08838},
  year={2024}
}

@article{banerjee2026intrinsic,
  title={Intrinsic geometry-inspired dependent toroidal distribution: application to regression model for astigmatism data},
  author={Banerjee, Buddhananda and Biswas, Surojit},
  journal={Computational Statistics},
  volume={41},
  number={2},
  pages={37},
  year={2026},
  publisher={Springer}
}

@article{giannakis2019data,
  title={Data-driven spectral decomposition and forecasting of ergodic dynamical systems},
  author={Giannakis, Dimitrios},
  journal={Applied and Computational Harmonic Analysis},
  volume={47},
  number={2},
  pages={338--396},
  year={2019},
  publisher={Elsevier}
}

@article{Eltzner2018torus,
    author= {Benjamin Eltzner and Stephan Huckemann and Kanti V. Mardia},
    title={Torus principal component analysis with applications to RNA structure}, 
    journal={The Annals of Applied Statistics},
    volume={12},
    number={2},
    pages={1332-1359},
    year={2018}
}

@inproceedings{sato2023online,
  title={Online Estimation of Self-Body Deflection With Various Sensor Data Based on Directional Statistics},
  author={Sato, Hiroya and Kawaharazuka, Kento and Makabe, Tasuku and Okada, Kei and Inaba, Masayuki},
  booktitle={2023 IEEE/SICE International Symposium on System Integration (SII)},
  pages={1--8},
  year={2023},
  organization={IEEE}
}

@article{pewseygarcia2021,
  author  = {Pewsey, Arthur and Garc{\'i}a-Portugu{\'e}s, Eduardo},
  title   = {Recent Advances in Directional Statistics},
  journal = {TEST},
  volume  = {30},
  number  = {1},
  pages   = {1--58},
  year    = {2021},
  doi     = {10.1007/s11749-021-00759-x}
}

@article{rousseeuw1987silhouettes,
	author  = {Rousseeuw, Peter J.},
	title   = {Silhouettes: A Graphical Aid to the Interpretation and
	Validation of Cluster Analysis},
	journal = {Journal of Computational and Applied Mathematics},
	volume  = {20},
	pages   = {53--65},
	year    = {1987},
	doi     = {10.1016/0377-0427(87)90125-7}
}

@article{calinski1974dendrite,
	author  = {Cali{\'n}ski, Tadeusz and Harabasz, Joachim},
	title   = {A Dendrite Method for Cluster Analysis},
	journal = {Communications in Statistics},
	volume  = {3},
	number  = {1},
	pages   = {1--27},
	year    = {1974},
	doi     = {10.1080/03610927408827101}
}

@article{anderson2001new,
	author  = {Anderson, Marti J.},
	title   = {A New Method for Non-Parametric Multivariate Analysis
	of Variance},
	journal = {Austral Ecology},
	volume  = {26},
	number  = {1},
	pages   = {32--46},
	year    = {2001},
	doi     = {10.1111/j.1442-9993.2001.01070.pp.x}
}

@article{mardia2007,
  author  = {Mardia, Kanti V. and Taylor, Charles C. and Subramaniam, Ganesh K.},
  title   = {Protein Bioinformatics and Mixtures of Bivariate von Mises Distributions for Angular Data},
  journal = {Biometrics},
  volume  = {63},
  number  = {2},
  pages   = {505--512},
  year    = {2007},
  doi     = {10.1111/j.1541-0420.2006.00682.x}
}

@article{mardia2012,
  author  = {Mardia, Kanti V. and Kent, John T. and Zhang, Zhengzheng and Taylor, Charles C. and Hamelryck, Thomas},
  title   = {Mixtures of Concentrated Multivariate Sine Distributions with Applications to Bioinformatics},
  journal = {Journal of Applied Statistics},
  volume  = {39},
  number  = {11},
  pages   = {2475--2492},
  year    = {2012},
  doi     = {10.1080/02664763.2012.719221}
}

@article{jung2021clustering,
  author  = {Jung, Sungkyu and Park, Kiho and Kim, Byungwon},
  title   = {Clustering on the Torus by Conformal Prediction},
  journal = {The Annals of Applied Statistics},
  volume  = {15},
  number  = {4},
  pages   = {1583--1603},
  year    = {2021},
  doi     = {10.1214/21-AOAS1459}
}

@manual{hongjung2022,
  author = {Jung, Sungkyu and Hong, Seungki and Park, Kiho and Kim, Byungwon},
  title  = {{ClusTorus}: Prediction and Clustering on the Torus by Conformal Prediction},
  note   = {R package version 0.2.2},
  year   = {2022},
  doi    = {10.32614/CRAN.package.ClusTorus}
}

@article{mardia2022pca,
  author  = {Mardia, Kanti V. and Wiechers, Henrik and Eltzner, Benjamin and Huckemann, Stephan F.},
  title   = {Principal Component Analysis and Clustering on Manifolds},
  journal = {Journal of Multivariate Analysis},
  volume  = {188},
  pages   = {104862},
  year    = {2022},
  doi     = {10.1016/j.jmva.2021.104862}
}

@article{wuyao2026,
  author  = {Wu, Menghao and Yao, Zhigang},
  title   = {A Principal Submanifold-Based Approach for Clustering and Multiscale RNA Correction},
  journal = {The Annals of Applied Statistics},
  volume  = {20},
  number  = {2},
  year    = {2026},
  doi     = {10.1214/26-AOAS2160}
}

@article{pennec2006,
  author  = {Pennec, Xavier},
  title   = {Intrinsic Statistics on Riemannian Manifolds: Basic Tools for Geometric Measurements},
  journal = {Journal of Mathematical Imaging and Vision},
  volume  = {25},
  number  = {1},
  pages   = {127--154},
  year    = {2006},
  doi     = {10.1007/s10851-006-6228-4}
}

@article{subbarao2009,
  author  = {Subbarao, Raghav and Meer, Peter},
  title   = {Nonlinear Mean Shift over Riemannian Manifolds},
  journal = {International Journal of Computer Vision},
  volume  = {84},
  number  = {1},
  pages   = {1--20},
  year    = {2009},
  doi     = {10.1007/s11263-008-0195-8}
}

@article{wang2013,
  author  = {Wang, Yang and Huang, Xiaodi and Wu, Lin},
  title   = {Clustering via Geometric Median Shift over Riemannian Manifolds},
  journal = {Information Sciences},
  volume  = {220},
  pages   = {292--305},
  year    = {2013},
  doi     = {10.1016/j.ins.2012.07.009}
}

@book{srivastava2016functional,
  author    = {Srivastava, Anuj and Klassen, Eric P.},
  title     = {Functional and Shape Data Analysis},
  series    = {Springer Series in Statistics},
  publisher = {Springer},
  address   = {New York},
  year      = {2016},
  doi       = {10.1007/978-1-4939-4020-2}
}

@article{llibre2024,
  author  = {Llibre, Jaume and Valls, Cl{\`a}udia},
  title   = {Darboux Theory of Integrability in $T^n$},
  journal = {European Journal of Mathematics},
  volume  = {10},
  pages   = {26},
  year    = {2024},
  doi     = {10.1007/s40879-024-00737-1}
}

@article{aaltoverma2019,
  author  = {Aalto, Max and Verma, Nakul},
  title   = {Metric Learning on Manifolds},
  journal = {arXiv preprint arXiv:1902.01738},
  year    = {2019}
}

@inproceedings{dutta2020unsupervised,
  author    = {Dutta, Ujjal K. and Harandi, Mehrtash and Sekhar, C. Chandra},
  title     = {Unsupervised Metric Learning with Synthetic Examples},
  booktitle = {Proceedings of the AAAI Conference on Artificial Intelligence},
  volume    = {34},
  pages     = {3834--3841},
  year      = {2020},
  doi       = {10.1609/aaai.v34i04.5795}
}

@article{tibshiraniwalther2005,
  author  = {Tibshirani, Robert and Walther, Guenther},
  title   = {Cluster Validation by Prediction Strength},
  journal = {Journal of Computational and Graphical Statistics},
  volume  = {14},
  number  = {3},
  pages   = {511--528},
  year    = {2005},
  doi     = {10.1198/106186005X59243}
}

@article{tibshirani2001gap,
  author  = {Tibshirani, Robert and Walther, Guenther and Hastie, Trevor},
  title   = {Estimating the Number of Clusters in a Data Set via the Gap Statistic},
  journal = {Journal of the Royal Statistical Society: Series B},
  volume  = {63},
  number  = {2},
  pages   = {411--423},
  year    = {2001},
  doi     = {10.1111/1467-9868.00293}
}

@inproceedings{benhur2002stability,
  author    = {Ben-Hur, Asa and Elisseeff, Andr{\'e} and Guyon, Isabelle},
  title     = {A Stability Based Method for Discovering Structure in Clustered Data},
  booktitle = {Pacific Symposium on Biocomputing},
  volume    = {7},
  pages     = {6--17},
  year      = {2002},
  doi       = {10.1142/9789812799623_0002}
}

\clearpage

\appendix

% ============================================================
% APPENDIX A: SUPPLEMENTARY NUMERICAL VALIDATION
% ============================================================

\clearpage

\appendix

\section{Additional geometry and proofs}
\label{app:geometry}

\subsection{Coordinatewise standardization of the recursive metric}
\label{app:metric-normalization}
The normalized family in the main text is a statistical, coordinatewise-standardized geometry inspired by the recursive Euclidean embedding; it is not obtained by dividing the complete induced metric by a single global constant. Suppose the unnormalized diagonal metric is
$$
\dd s^2=\sum_{k=1}^N H_k(\theta)^2\dd\theta_k^2,
$$
with orthogonal scale factors
$$
H_1=B_1,\quad H_k=B_k+H_{k-1}\cos\theta_{k-1},\quad k=2,\ldots,N,
$$
where $B_k>0$ is the baseline scale associated with coordinate $k$. We standardize each scale factor by its own baseline, $h_k=H_k/B_k$. Then
$$
h_1=1,\quad h_k=1+\lambda_k h_{k-1}\cos\theta_{k-1},\quad \lambda_k=B_{k-1}/B_k.
$$
This produces the dimensionless statistical metric
$$
\dd\widetilde s^2=\sum_{k=1}^N h_k(\theta)^2\dd\theta_k^2.
$$
The paper uses the parsimonious specialization $\lambda_2=\cdots=\lambda_N=\lambda$, giving
\begin{equation}
\label{eq:supp-common-lambda}
h_1=1,\quad h_k=1+\lambda h_{k-1}\cos\theta_{k-1},\quad k=2,\ldots,N.
\end{equation}
For $N=2$ this is $h_2=1+\lambda\cos\theta_1$ and hence recovers \eqref{eq:t2metric}; for $\lambda=0$, every $h_k\equiv1$ and the flat product-torus metric is recovered. The coordinatewise standardization deliberately removes baseline units separately while retaining the recursive position-dependent modulation.

\subsection{Angular-origin dependence}
Let $T_\alpha([\theta])=[\theta+\alpha]$ be a toroidal phase relabelling. If the functional form \eqref{eq:hk} is held fixed, replacing $\theta$ by $\theta+\alpha$ generally changes the coefficients because $\cos\theta_j$ becomes $\cos(\theta_j+\alpha_j)$. Thus the construction is not invariant to an arbitrary change of angular origin when the metric is held fixed. If the metric is transformed simultaneously, $g^{(\alpha)}=(T_\alpha^{-1})^*g$, then the corresponding path energies and CTE dissimilarities are preserved by the same pullback argument used for coordinate relabelling. Hence, reference phases, like coordinate order, are part of the prespecified geometric convention.

    \subsection{Proof of Theorem \ref{thm:recursive-positivity}}
    \label{app:recursive-positivity}
    
    \begin{proof}
        We first establish pointwise upper and lower bounds for the recursive
        coefficients. Since $    h_1(\theta;\lambda)=1,$
        we have $    |h_1(\theta;\lambda)|=b_1(\lambda)=1.$
        Assume for some $k\ge2$ that $    |h_{k-1}(\theta;\lambda)|
        \le
        b_{k-1}(\lambda)$
        for all $\theta\in\mathbb T^N$. Using
        $|\cos\theta_{k-1}|\le1$ and $\lambda\ge0$,
        \begin{align}
            |h_k(\theta;\lambda)|  \nonumber
            &= \left| 1 + \lambda h_{k-1}(\theta;\lambda)\cos\theta_{k-1} \right| \\ \nonumber
            &\le 1 + \lambda |h_{k-1}(\theta;\lambda)| \\ \nonumber
            &\le 1 + \lambda b_{k-1}(\lambda). \nonumber
        \end{align}

        Because
        $$
        1+\lambda b_{k-1}(\lambda)
        =
        1+\lambda\sum_{j=0}^{k-2}\lambda^j
        =
        \sum_{j=0}^{k-1}\lambda^j
        =
        b_k(\lambda),
        $$
        induction gives $    |h_k(\theta;\lambda)|
        \le
        b_k(\lambda),
        \quad
        k=1,\ldots,N.$
        The same bound yields the lower estimate. For $k\ge2$,
        %        $$
        %        h_k(\theta;\lambda)
        %        =
        %        1+\lambda h_{k-1}(\theta;\lambda)\cos\theta_{k-1}
        %        \ge
        %        1-\lambda |h_{k-1}(\theta;\lambda)|
        %        \ge
        %        1-\lambda b_{k-1}(\lambda).
        %        $$
        \begin{align}
            h_k(\theta;\lambda) \nonumber
            &= 1 + \lambda h_{k-1}(\theta;\lambda)\cos\theta_{k-1} \\\nonumber
            &\ge 1 - \lambda |h_{k-1}(\theta;\lambda)| \\\nonumber
            &\ge 1 - \lambda b_{k-1}(\lambda).\nonumber
        \end{align}
        Since  $        1-\lambda b_{k-1}(\lambda)
        =
        1-\sum_{j=1}^{k-1}\lambda^j
        =
        a_k(\lambda),$ then         for every $k$
        we obtain $    a_k(\lambda)
        \le
        h_k(\theta;\lambda)
        \le
        b_k(\lambda).$
        %        \begin{equation}
            %            \label{eq:proof-lower}
            %            a_k(\lambda)
            %            \le
            %            h_k(\theta;\lambda)
            %            \le
            %            b_k(\lambda).
            %        \end{equation}
        
        It remains to make these inequalities uniform over
        $\lambda\in[0,\lambda_0]$ and $k\le N$. Since all powers of
        $\lambda$ are nonnegative and nondecreasing on this interval, $\displaystyle         \sum_{j=1}^{k-1}\lambda^j
        \le
        \sum_{j=1}^{N-1}\lambda_0^j.$
        Therefore,
        $$
        a_k(\lambda)
        =
        1-\sum_{j=1}^{k-1}\lambda^j
        \ge
        1-\sum_{j=1}^{N-1}\lambda_0^j
        =
        \underline h_{N,\lambda_0}
        >
        0,
        $$
        where strict positivity follows from
        \eqref{eq:uniform-positivity-condition}. Likewise,
        $$
        b_k(\lambda)
        =
        \sum_{j=0}^{k-1}\lambda^j
        \le
        \sum_{j=0}^{N-1}\lambda_0^j
        =
        \overline h_{N,\lambda_0}.
        $$
        This proves \eqref{eq:recursive-uniform-bounds}.
        Finally, because $g_\lambda(\theta)$ is diagonal,
        $$
        g_{\lambda,\theta}(v,v)
        =
        \sum_{k=1}^{N}
        h_k(\theta;\lambda)^2v_k^2.
        $$
        Applying \eqref{eq:recursive-uniform-bounds} term by term gives
        $$
        \underline h_{N,\lambda_0}^{\,2}
        \sum_{k=1}^{N}v_k^2
        \le
        \sum_{k=1}^{N}
        h_k(\theta;\lambda)^2v_k^2
        \le
        \overline h_{N,\lambda_0}^{\,2}
        \sum_{k=1}^{N}v_k^2,
        $$
        which is exactly \eqref{eq:uniform-ellipticity}. Thus every
        $g_\lambda(\theta)$ is positive definite, with a strictly positive
        lower eigenvalue bound that is uniform in both $\theta$ and
        $\lambda\in[0,\lambda_0]$.
    \end{proof}

    \subsection{Proof of Theorem \ref{thm:coordinate_relabelling}}
    \label{app}
    Let $    x=[\theta],
    \quad
    y=[\eta],$
    and denote their coordinatewise principal angular difference by $    \delta=\wrap(\eta-\theta).$
    
    Under the relabelling $F_\pi$, the corresponding representatives are $    \theta^{(\pi)}=P_\pi\theta,
    \quad
    \eta^{(\pi)}=P_\pi\eta.$
    
    Since $\wrap(\cdot)$ acts coordinatewise, it commutes with coordinate
    permutation. Hence
    $$
    \delta^{(\pi)}
    =
    \wrap\!\left(\eta^{(\pi)}-\theta^{(\pi)}\right)
    =
    \wrap\!\left(P_\pi(\eta-\theta)\right)
    $$
    
    $$    =
    P_\pi\wrap(\eta-\theta)
    =
    P_\pi\delta.$$
    
    Now fix an arbitrary winding vector $m\in\Z^N$ and define its $    m^{(\pi)}=P_\pi m.$
    relabelled version by
    
    Because $P_\pi$ merely permutes integer coordinates, $    m\longmapsto m^{(\pi)}$
    is a bijection from $\Z^N$ onto itself. Moreover,
    $$
    \begin{aligned}
        \Delta^{(m^{(\pi)})}
        &=
        \delta^{(\pi)}+2\pi m^{(\pi)}
        \\
        &=
        P_\pi\delta+2\pi P_\pi m
        \\
        &=
        P_\pi(\delta+2\pi m)
        \\
        &=
        P_\pi\Delta^{(m)}.
    \end{aligned}
    $$
    
    Therefore the coordinate-linear lifted path associated with
    $m^{(\pi)}$ in the relabelled coordinates is
    $$
    \begin{aligned}
        \gamma_{m^{(\pi)}}^{(\pi)}(t)
        &=
        \theta^{(\pi)}
        +t\Delta^{(m^{(\pi)})}
        \\
        &=
        P_\pi\theta+tP_\pi\Delta^{(m)}
        \\
        &=
        P_\pi\bigl(\theta+t\Delta^{(m)}\bigr)
        \\
        &=
        F_\pi\bigl(\gamma_m(t)\bigr).
    \end{aligned}
    $$
    Differentiating gives $    \dot\gamma_{m^{(\pi)}}^{(\pi)}(t)
    =
    dF_\pi\!\left(\dot\gamma_m(t)\right).$
    By definition,  $    g^{(\pi)}=(F_\pi^{-1})^*g.$
    Thus $F_\pi:(\T^N,g)\to(\T^N,g^{(\pi)})$ is an isometry, so for every
    $p\in\T^N$ and every tangent vector $v\in T_p\T^N$,
    $$
    g^{(\pi)}_{F_\pi(p)}
    \left(
    dF_\pi(v),dF_\pi(v)
    \right)
    =
    g_p(v,v).
    $$
    Taking $    p=\gamma_m(t),
    \quad
    v=\dot\gamma_m(t),$
    we obtain
    $$
    g^{(\pi)}_{\gamma_{m^{(\pi)}}^{(\pi)}(t)}
    \left(
    \dot\gamma_{m^{(\pi)}}^{(\pi)}(t),
    \dot\gamma_{m^{(\pi)}}^{(\pi)}(t)
    \right)
    =
    g_{\gamma_m(t)}
    \left(
    \dot\gamma_m(t),\dot\gamma_m(t)
    \right).
    $$
    Integrating over $t\in[0,1]$ yields \\ $    E_{m^{(\pi)}}^{\,g^{(\pi)}}
    \bigl(F_\pi(x),F_\pi(y)\bigr)
    =
    E_m^{\,g}(x,y).$
    Since $m\mapsto P_\pi m$ is a bijection of $\Z^N$,
    $$
    \begin{aligned}
        \E^{\,g^{(\pi)}}\bigl(F_\pi(x),F_\pi(y)\bigr)
        &=
        \min_{m'\in\Z^N}
        E_{m'}^{\,g^{(\pi)}}\bigl(F_\pi(x),F_\pi(y)\bigr)
        \\
        &=
        \min_{m\in\Z^N}
        E_{P_\pi m}^{\,g^{(\pi)}}\bigl(F_\pi(x),F_\pi(y)\bigr)
        \\
        &=
        \min_{m\in\Z^N}
        E_m^{\,g}(x,y)
        \\
        &=
        \E^{\,g}(x,y).
    \end{aligned}
    $$
    Taking square roots proves
    $$
    \DE^{\,g^{(\pi)}}\bigl(F_\pi(x),F_\pi(y)\bigr)
    =
    \DE^{\,g}(x,y).
    $$

    \subsection{Proof of Theorem~\ref{thm:finite}}
Since $h_k(\vartheta)\ge a_k>0$ for every
$\vartheta\in\T^N$, we have along every lifted path
$$
h_k(\gamma_m(t))^2\ge a_k^2,
\quad t\in[0,1].
$$
Moreover, $\Delta_k^{(m)}=\delta_k+2\pi m_k$ is constant
along the path. Hence
\begin{align}
E_m(x,y)
&=
\int_0^1\sum_{k=1}^N
h_k(\gamma_m(t))^2
\bigl(\Delta_k^{(m)}\bigr)^2\,dt\nonumber\\
&\ge
\int_0^1\sum_{k=1}^N
a_k^2
\bigl(\Delta_k^{(m)}\bigr)^2\,dt \nonumber\\
&=
\sum_{k=1}^N
a_k^2
\bigl(\delta_k+2\pi m_k\bigr)^2.
\label{eq:app-lower}
\end{align}
    Because the zero winding is feasible, \\
    $\E(x,y)=\min_m E_m(x,y)\le E_0(x,y)=U$. Hence every minimizing
    winding $m^\star$ satisfies $E_{m^\star}\le U$, and
    \eqref{eq:app-lower} gives
    $$
    \sum_{k=1}^N a_k^2
    \bigl(\delta_k+2\pi m_k^\star\bigr)^2\le U.
    $$
    This proves \eqref{eq:ellipsoid}.
    
    For each coordinate separately,
    $$
    -\frac{\sqrt U}{a_k}
    \le
    \delta_k+2\pi m_k^\star
    \le
    \frac{\sqrt U}{a_k}.
    $$
    Solving for the integer $m_k^\star$ yields exactly
    $$
    \left\lceil
    \frac{-\sqrt U/a_k-\delta_k}{2\pi}
    \right\rceil
    \le m_k^\star\le
    \left\lfloor
    \frac{\sqrt U/a_k-\delta_k}{2\pi}
    \right\rfloor,
    $$
    which is \eqref{eq:windinterval}. Define the admissible winding set
$$
\mathcal M_U(x,y)
=
\left\{
m\in\mathbb Z^N:
\sum_{k=1}^N
a_k^2
\bigl(\delta_k+2\pi m_k\bigr)^2
\le U
\right\}.
$$
By the preceding lower bound, every minimizing winding vector belongs to
$\mathcal M_U(x,y)$. Now let $m\in\mathcal M_U(x,y)$. Since every term in the defining sum
is nonnegative, for each $k$,
$
a_k^2\bigl(\delta_k+2\pi m_k\bigr)^2
\le U.
$
Hence
$
|\delta_k+2\pi m_k|
\le
\frac{\sqrt U}{a_k},
$
which is equivalent to
$
\ell_k(U)\le m_k\le r_k(U).
$
Therefore,
$$
\mathcal M_U(x,y)
\subseteq
\prod_{k=1}^N
\{\ell_k(U),\ldots,r_k(U)\}.
$$
In particular, $\mathcal M_U(x,y)$ is finite. Since the $k$th factor
contains $r_k(U)-\ell_k(U)+1$ integers, taking cardinalities gives
$$
|\mathcal M_U(x,y)|
\le
\prod_{k=1}^N
\bigl(r_k(U)-\ell_k(U)+1\bigr).
$$
A simpler symmetric bound follows from $|\delta_k|\le\pi$. Indeed,
$$
2\pi|m_k|
=
|2\pi m_k|
\le
|\delta_k+2\pi m_k|+|\delta_k|
\le
\frac{\sqrt U}{a_k}+\pi.
$$
Thus
$
|m_k|
\le
\frac{1}{2\pi}
\left(
\frac{\sqrt U}{a_k}+\pi
\right).
$
Since $m_k$ is integer, this implies
$
|m_k|\le M_k(U),
$
where $M_k(U)$ is defined in \eqref{eq:mkbound}. Consequently,
$$
\mathcal M_U(x,y)
\subseteq
\prod_{k=1}^N
\{-M_k(U),\ldots,M_k(U)\}.
$$
The $k$th factor contains $2M_k(U)+1$ integers, and hence
$$
|\mathcal M_U(x,y)|
\le
\prod_{k=1}^N
\bigl(2M_k(U)+1\bigr).
$$
Together, these establish the two bounds in
\eqref{eq:candidatebound}.
    
    We next exploit the recursive form \eqref{eq:hk}. For fixed
    $(m_1,\ldots,m_{N-1})$, every coefficient
    $h_k(\gamma_m(t))$ is independent of $m_N$, by recursion, $h_k$ depends
    only on the preceding coordinates $\theta_1,\ldots,\theta_{k-1}$, and in
    particular, no coefficient depends on $\theta_N$. Hence, the energy can be
    written as
    \begin{eqnarray}
E_m(x,y)
& = & A(m_1,\ldots,m_{N-1})\nonumber\\
&+& c(m_1,\ldots,m_{N-1})
\bigl(\delta_N+2\pi m_N\bigr)^2.\nonumber
\end{eqnarray}
    where
    $ \displaystyle
    c(m_1,\ldots,m_{N-1})
    =\int_0^1 h_N(\gamma_m(t))^2\,\mathrm dt>0
    $
    is independent of $m_N$. Because $\delta_N\in(-\pi,\pi]$, the choice
    $m_N=0$ minimizes $|\delta_N+2\pi m_N|$. If $\delta_N=\pi$, then
    $m_N=-1$ ties with $m_N=0$; choosing $m_N=0$ fixes a canonical
    representative without changing the minimum. Thus the winding search may be
    restricted to $\mathcal M_U^{\mathrm{can}}(x,y)$. Applying the already proved
    coordinatewise bounds only to $k=1,\ldots,N-1$ gives
    \eqref{eq:canonicalcount}.
    
    \subsection{Proof of Corollary~\ref{cor:uniform-winding}}
    For the constant-$\lambda$ recursive family, the bounds from the preceding
    subsection give
    $$
    0<a_k\le h_k(\vartheta)\le b_k,
    \quad
    b_k=\sum_{j=0}^{k-1}\lambda^j,
    \quad
    a_k=1-\sum_{j=1}^{k-1}\lambda^j,
    $$
    with $a_1=b_1=1$. Along the principal lift, $|\delta_k|\le\pi$, so
    \begin{align*}
        U
        &=E_0(x,y)\\
        &=\int_0^1\sum_{k=1}^{N}
        h_k(\gamma_0(t))^2\delta_k^2\,\mathrm dt \le \sum_{k=1}^{N}b_k^2\delta_k^2\\
        &\le \pi^2\sum_{k=1}^{N}b_k^2
        =: \overline U_{N,\lambda}.
    \end{align*}
    Therefore replacing $U$ by $\overline U_{N,\lambda}$ in the preceding
    coordinate bounds yields a deterministic pair-uniform candidate box that
    contains every canonical minimizing winding. This proves
    \eqref{eq:uglobal}-\eqref{eq:cglobal}.
    
    \subsection{Proof of Proposition~\ref{prop:cte-complexity}}
    Consider $Q$-node Gauss-Legendre evaluation of one candidate
    energy. At a fixed quadrature node, the recursion
    $h_1=1$, $h_k=1+\lambda h_{k-1}\cos\theta_{k-1}$ computes all $N$
    metric coefficients in $O(N)$ operations, after which the quadratic energy
    sum is also $O(N)$. Thus one candidate costs $O(QN)$ operations. Evaluating
    at most $\overline C_{N,\lambda}$ canonical candidates gives
    $O(QN\overline C_{N,\lambda})$ worst-case arithmetic cost per pair;
    recursive ellipsoidal pruning can only reduce this count. A dense symmetric
    $n\times n$ matrix contains $n(n-1)/2$ nontrivial pairs, giving
    $O(n^2QN\overline C_{N,\lambda})$ arithmetic cost.
    
    To see the uniform interior statement, fix $\lambda_0$ such that
    $\sum_{j=1}^{N-1}\lambda_0^j<1$ and restrict $0\le\lambda\le\lambda_0$.
    Then
    $$
    a_k(\lambda)\ge
    1-\sum_{j=1}^{N-1}\lambda_0^j>0,
    \quad
    b_k(\lambda)\le
    \sum_{j=0}^{N-1}\lambda_0^j<\infty
    $$
    for every $k\le N$. Hence $\overline U_{N,\lambda}$ and every
    $\overline M_k$ are bounded uniformly on the interval, and therefore so is
    $\overline C_{N,\lambda}$. This completes the proof.

    \subsection{Proof of Theorem~\ref{thm:cte-properties}}
    \label{app:cte-properties-proof}
    
    \begin{proof}
        Let
        $
        \delta^{xy}
        =
        \wrap(\eta-\theta)
        $
        and, for $m\in\Z^N$, write
        $
        \Delta_{xy}^{(m)}
        =
        \delta^{xy}+2\pi m.
        $
        By the finite-winding theorem, the minimum defining
        $\E_\lambda(x,y)$ is attained over a finite candidate set, so
        $\DEpar{\lambda}$ is well defined.
        
       \paragraph{Lift invariance.}
Replace the representatives by
$$
\theta'=\theta+2\pi p,
\quad
\eta'=\eta+2\pi q,
\quad
p,q\in\Z^N.
$$
Then
$
\eta'-\theta'
=
(\eta-\theta)+2\pi(q-p).
$
Since $q-p\in\Z^N$ and the wrapping map is $2\pi$-periodic
coordinatewise,
$$
\wrap(u+2\pi r)=\wrap(u),
\quad r\in\Z,
$$
we obtain
$
\wrap(\eta'-\theta')
=
\wrap(\eta-\theta)
=
\delta^{xy}.
$
Thus the wrapped displacement is independent of the chosen lifts
$\theta$ and $\eta$.
        % For a fixed winding vector $m$, the corresponding lifted path is
        % shifted by the constant vector $2\pi p$. Since the recursive metric
        % coefficients are $2\pi$-periodic in every angular coordinate, this
        % shift leaves the path energy unchanged. Hence every candidate energy,
        % and therefore its minimum, is independent of the chosen lifts.
        
        \paragraph{Symmetry:}
        Let
        $
        \delta^{yx}
        =
        \wrap(\theta-\eta).
        $
        Because both $\delta^{yx}$ and $-\delta^{xy}$ represent the same
        element of $\T^N$, there exists $q\in\Z^N$ such that
        $$
        \delta^{yx}
        =
        -\delta^{xy}+2\pi q.
        $$
        For any $m\in\Z^N$, define
        $
        m'=-m-q.
        $
        Then
        $$
        \Delta_{yx}^{(m')}
        =
        \delta^{yx}+2\pi m'
        =
        -\delta^{xy}-2\pi m
        =
        -\Delta_{xy}^{(m)}.
        $$
        The map $m\mapsto m'$ is a bijection of $\Z^N$.
        
        The reverse of the torus path associated with $m$ is again
        coordinate-linear and has increment $-\Delta_{xy}^{(m)}$.
        Its lift differs from the candidate path associated with $m'$ only by
        a constant vector in $2\pi\Z^N$. Periodicity of the metric and the
        change of variable $t\mapsto1-t$ therefore give
        $
        E_{\lambda,m}^{xy}
        =
        E_{\lambda,m'}^{yx}.
        $
        Taking minima over the bijectively corresponding winding vectors gives
        $
        \E_\lambda(x,y)=\E_\lambda(y,x),
        $
        and hence
        $$
        \DEpar{\lambda}(x,y)=\DEpar{\lambda}(y,x).
        $$
        
        \paragraph{Non-negativity and identity of indiscernibles:}
        Every candidate energy is nonnegative because, $g_\lambda$ is positive
        definite. If $x=y$, choosing the zero winding gives zero energy, so
        $\DEpar{\lambda}(x,x)=0$.
        
        Conversely, suppose $\DEpar{\lambda}(x,y)=0$. Since the minimum is
        attained, there exists a minimizing winding $m^\star$ such that
        $$
        0
        =
        \int_0^1
        \sum_{k=1}^N
        h_k(\gamma_{m^\star}(t);\lambda)^2
        \bigl(\Delta_{xy,k}^{(m^\star)}\bigr)^2
        \,\dd t .
        $$
        Strict positivity of every metric coefficient implies
        $$
        \Delta_{xy,k}^{(m^\star)}=0,
        \quad
        k=1,\ldots,N.
        $$
        Thus $\eta-\theta\in2\pi\Z^N$, so $x=y$ in $\T^N$.
        
   \paragraph{Comparison with the geodesic distance.}
For an absolutely continuous path
$\gamma:[0,1]\to\T^N$, define its Riemannian length by
$$
L_{g_\lambda}(\gamma)
=
\int_0^1
\|\dot\gamma(t)\|_{g_\lambda}\,\dd t,
$$
where
$
\|\dot\gamma(t)\|_{g_\lambda}
=
\sqrt{
g_{\lambda,\gamma(t)}
\bigl(\dot\gamma(t),\dot\gamma(t)\bigr)
}.
$
Its Riemannian energy is
$
E_{g_\lambda}(\gamma)
=
\int_0^1
\|\dot\gamma(t)\|_{g_\lambda}^2\,\dd t.
$
By the Cauchy-Schwarz inequality,
$$
\begin{aligned}
L_{g_\lambda}(\gamma)^2
&=
\left(
\int_0^1
\|\dot\gamma(t)\|_{g_\lambda}\,\dd t
\right)^2 \\
&\le
\left(\int_0^1 1\,\dd t\right)
\left(
\int_0^1
\|\dot\gamma(t)\|_{g_\lambda}^2\,\dd t
\right) \\
&=
E_{g_\lambda}(\gamma).
\end{aligned}
$$
        Therefore every CTE candidate path satisfies
        $$
        d_{g_\lambda}(x,y)
        \le
        L_{g_\lambda}(\gamma_m)
        \le
        \sqrt{E_{\lambda,m}(x,y)}.
        $$
        Taking the minimum over all winding classes yields
        $$
        d_{g_\lambda}(x,y)
        \le
        \DEpar{\lambda}(x,y).
        $$
        
        Now assume $a\le h_k(\theta;\lambda)\le b$. Pointwise domination of
        the quadratic forms gives $    a^2g_0\preceq g_\lambda\preceq b^2g_0.$
        Hence every path satisfies  $        L_{g_\lambda}(\gamma)\ge aL_{g_0}(\gamma),$
        and taking infima gives
        $$
        d_{g_\lambda}(x,y)\ge a\,d_0(x,y).
        $$
        On the other hand, the principal coordinate-linear lift $m=0$ has
        increment $\delta^{xy}$, so
        $$
        \E_\lambda(x,y)
        \le
        E_{\lambda,0}(x,y)
        \le
        b^2
        \sum_{k=1}^N
        (\delta_k^{xy})^2
        =
        b^2d_0(x,y)^2.
        $$
        Taking square roots proves
        $$
        a\,d_0(x,y)
        \le
        d_{g_\lambda}(x,y)
        \le
        \DEpar{\lambda}(x,y)
        \le
        b\,d_0(x,y).
        $$
        
        \paragraph{Flat endpoint:}
        At $\lambda=0$, $    h_k(\theta;0)=1,
        \quad k=1,\ldots,N.$
        Thus
        $$
        E_{0,m}(x,y)
        =
        \sum_{k=1}^N
        \bigl(\delta_k^{xy}+2\pi m_k\bigr)^2.
        $$
        The minimization separates coordinatewise. Since
        $\delta_k^{xy}\in(-\pi,\pi]$, the principal choice $m_k=0$ is a
        minimizer in each coordinate, if $\delta_k^{xy}=\pi$, the choice
        $m_k=-1$ ties with $m_k=0$ and gives the same squared displacement.
        Therefore
        $$
        \E_0(x,y)
        =
        \sum_{k=1}^N(\delta_k^{xy})^2,
        $$
        and
        $$
        \DEpar{0}(x,y)=d_0(x,y)=d_{g_0}(x,y).
        $$
    \end{proof}
    
    \subsection{Proof of Theorem~\ref{thm:lambda-stability}}
    \label{app:lambda-stability-proof}
    
    \begin{proof}
        For $k=1,\ldots,N$, define,
        $
        A_k=
        1-\sum_{j=1}^{k-1}\lambda_0^j,
        \quad
        B_k
        =
        \sum_{j=0}^{k-1}\lambda_0^j,
        $
        and
        $
        D_1=0,
        \quad
        D_k
        =
        \sum_{j=1}^{k-1}j\lambda_0^{\,j-1},
        \quad k\ge2.
        $
        By Theorem~\ref{thm:recursive-positivity}, $A_k>0$ and
        $
        A_k\le h_k(\theta;\nu)\le B_k
        $
        uniformly over $\theta\in\T^N$ and $\nu\in[0,\lambda_0]$. Now, since $h_1\equiv1$, $\partial_\nu h_1=0$. For $k\ge2$,
        $$
        \partial_\nu h_k
        =
        h_{k-1}\cos\theta_{k-1}
        +
        \nu(\partial_\nu h_{k-1})\cos\theta_{k-1}.
        $$

Using $|\cos\theta_{k-1}|\le1$, the induction hypothesis gives
$$
|\partial_\nu h_k|
\le
B_{k-1}+\lambda_0D_{k-1}.
$$
Moreover,
$$
\begin{aligned}
B_{k-1}+\lambda_0D_{k-1}
&=
\sum_{j=0}^{k-2}\lambda_0^j
+
\sum_{j=1}^{k-2}j\lambda_0^j\\
&=
1+2\lambda_0+3\lambda_0^2+\cdots
+(k-1)\lambda_0^{k-2}\\
&=
\sum_{j=1}^{k-1}j\lambda_0^{j-1}
=
D_k.
\end{aligned}
$$
Hence $|\partial_\nu h_k(\theta;\nu)|\le D_k$.

Fix a winding vector $m$ and write
$
\Delta_k=\delta_k+2\pi m_k.
$
The corresponding lifted coordinate-linear path
$\widetilde\gamma_m(t)=\theta+t\Delta$ does not depend on $\nu$.
Hence differentiation under the integral sign gives
$$
\begin{aligned}
\frac{\partial}{\partial\nu}E_{\nu,m}(x,y)
&=
2\int_0^1
\sum_{k=1}^N \Big[
h_k(\widetilde\gamma_m(t);\nu) \\
&\quad\quad
\partial_\nu h_k(\widetilde\gamma_m(t);\nu)
\Delta_k^2 \Big] dt.
\end{aligned}
$$
Using
$|h_k|\le B_k$ and $|\partial_\nu h_k|\le D_k$,
$$
\left|
\frac{\partial}{\partial\nu}E_{\nu,m}(x,y)
\right|
\le
2\sum_{k=1}^N B_kD_k\Delta_k^2.
$$
On the other hand, since $h_k\ge A_k$,
$$
E_{\nu,m}(x,y)
\ge
\sum_{k=1}^N A_k^2\Delta_k^2.
$$
Define
$
\displaystyle
C_{N,\lambda_0}
=
2\max_{1\le k\le N}
\frac{B_kD_k}{A_k^2}.
$
Then, for every $k$,
$$
2B_kD_k\Delta_k^2
\le
C_{N,\lambda_0}A_k^2\Delta_k^2.
$$
Therefore,
$$
\begin{aligned}
\left|
\frac{\partial}{\partial\nu}E_{\nu,m}(x,y)
\right|
&\le
C_{N,\lambda_0}
\sum_{k=1}^N A_k^2\Delta_k^2\\
&\le
C_{N,\lambda_0}E_{\nu,m}(x,y).
\end{aligned}
$$
If $\Delta\neq0$, then $E_{\nu,m}(x,y)>0$, and hence
$$
\left|
\frac{\partial_\nu E_{\nu,m}(x,y)}
     {E_{\nu,m}(x,y)}
\right|
\le
C_{N,\lambda_0}.
$$
Since
$$
\frac{\partial_\nu E_{\nu,m}}
     {E_{\nu,m}}
=
\partial_\nu\log E_{\nu,m},
$$
integration between $\lambda$ and $\mu$ gives
$$
-C_{N,\lambda_0}|\lambda-\mu|
\le
\log\frac{E_{\mu,m}(x,y)}
          {E_{\lambda,m}(x,y)}
\le
C_{N,\lambda_0}|\lambda-\mu|.
$$
Exponentiating yields
\begin{eqnarray}
 e^{-C_{N,\lambda_0}|\lambda-\mu|} E_{\lambda,m}(x,y) & \le &E_{\mu,m}(x,y) \nonumber\\
& \le & e^{C_{N,\lambda_0}|\lambda-\mu|} E_{\lambda,m}(x,y). \nonumber
\end{eqnarray}

If $\Delta=0$, then $E_{\nu,m}(x,y)=0$ for every $\nu$, so the
same inequality holds trivially.

Since the constants are independent of $m$, taking minima over
$m\in\Z^N$ gives
\begin{eqnarray}
e^{-C_{N,\lambda_0}|\lambda-\mu|}
\min_m E_{\lambda,m}(x,y) & \le & 
\min_m E_{\mu,m}(x,y) \nonumber\\
& \le & 
e^{C_{N,\lambda_0}|\lambda-\mu|}
\min_m E_{\lambda,m}(x,y), \nonumber
\end{eqnarray}

% $$
% e^{-C_{N,\lambda_0}|\lambda-\mu|}
% \min_m E_{\lambda,m}(x,y)
% \le
% \min_m E_{\mu,m}(x,y)
% \le
% e^{C_{N,\lambda_0}|\lambda-\mu|}
% \min_m E_{\lambda,m}(x,y),
% $$
which proves \eqref{eq:energy-multiplicative-stability}. Now  taking square roots gives
        \begin{eqnarray}
e^{-C_{N,\lambda_0}|\lambda-\mu|/2} \DEpar{\lambda}(x,y)
& \le & \DEpar{\mu}(x,y) \nonumber \\
& \le & e^{C_{N,\lambda_0}|\lambda-\mu|/2} \DEpar{\lambda}(x,y).\nonumber
\end{eqnarray}     
      Finally, using the principal winding $m=0$ and the uniform bound
$h_k\le B_N$, we obtain
$$
\begin{aligned}
\DEpar{\nu}(x,y)^2
&\le
E_{\nu,0}(x,y)\\
&=
\int_0^1
\sum_{k=1}^N
h_k(\gamma_0(t);\nu)^2\delta_k^2\,dt\\
&\le
B_N^2\sum_{k=1}^N\delta_k^2\\
&\le
B_N^2N\pi^2,
\end{aligned}
$$
because $|\delta_k|\le\pi$. Therefore,
$
\DEpar{\nu}(x,y)\le B_N\pi\sqrt N
$
        uniformly in $\nu,x,y$. Taking square roots in
\eqref{eq:energy-multiplicative-stability} gives, with
$s=|\lambda-\mu|$,
$$
e^{-C_{N,\lambda_0}s/2}\DEpar{\lambda}(x,y)
\le
\DEpar{\mu}(x,y)
\le
e^{C_{N,\lambda_0}s/2}\DEpar{\lambda}(x,y).
$$
If $\DEpar{\mu}(x,y)\ge\DEpar{\lambda}(x,y)$, then
$$
\DEpar{\mu}(x,y)-\DEpar{\lambda}(x,y)
\le
\left(e^{C_{N,\lambda_0}s/2}-1\right)
\DEpar{\lambda}(x,y).
$$
If the reverse inequality holds, the same argument with
$\lambda$ and $\mu$ interchanged gives
$$
\DEpar{\lambda}(x,y)-\DEpar{\mu}(x,y)
\le
\left(e^{C_{N,\lambda_0}s/2}-1\right)
\DEpar{\mu}(x,y).
$$
Since
$$
\DEpar{\nu}(x,y)\le B_N\pi\sqrt N
$$
uniformly for $\nu\in[0,\lambda_0]$, it follows that
$$
|\DEpar{\mu}(x,y)-\DEpar{\lambda}(x,y)|
\le
B_N\pi\sqrt N
\left(e^{C_{N,\lambda_0}s/2}-1\right).
$$
        Using $e^u-1\le ue^u$ and $s\le\lambda_0$ gives
        $$
        |\DEpar{\mu}(x,y)-\DEpar{\lambda}(x,y)|
        \le
        L_{N,\lambda_0}|\mu-\lambda|,
        $$
        where one valid constant is
        $$
        L_{N,\lambda_0}
        =
        \frac{C_{N,\lambda_0}}{2}
        e^{C_{N,\lambda_0}\lambda_0/2}
        B_N\pi\sqrt N.
        $$
        This proves \eqref{eq:distance-lipschitz} and the asserted continuity.
    \end{proof}
    
%     \subsection{Triangle-inequality counterexample}
% \label{app:triangle}
%     At $\lambda=0.55$ under the same normalized $\T^2$ metric, let $    x=(2.58345883,0.86911078)\\
%     y=(2.07120647,2.93318636),\\
%     z=(6.20770859,4.02709426).$
    
%     Exact finite-winding evaluation of the coordinate-linear energies gives
%     $$
%     \DE(x,y)=1.3963780876,\quad
%     \DE(y,z)=2.5575284824,
%     $$
%     and
%     $$
%     \DE(x,z)=4.5333022725.
%     $$
%     Hence
%     $$
%     \DE(x,z)-\DE(x,y)-\DE(y,z)=0.5793957024>0.
%     $$
%     This establishes that $\DE$ is not a metric in general.

\subsection{Triangle-inequality counterexample}
\label{app:triangle}

Consider the normalized $\T^2$ geometry with $\lambda=0.55$ and the
three points, $x = (2.5835,\;0.8691),~~
y = (2.0712,\;2.9332),~~
z = (6.2077,\;4.0271).$
The certified finite-winding search gives the CTE dissimilarities
$
\DE(x,y)=1.3964,
\quad
\DE(y,z)=2.5575,
$
and
$
\DE(x,z)=4.5333.
$
Therefore,
$$
\DE(x,z)
-
\DE(x,y)
-
\DE(y,z)
=
0.5794>0.
$$
equivalently,
$$
\DE(x,z)
>
\DE(x,y)+\DE(y,z),
$$
which violates the triangle inequality. Hence $\DE$ is, in general,
a dissimilarity rather than a metric.

\section{CEHS-HM selection details}
\label{app:cehs}

\paragraph{CEHS-HM geometry-selection algorithm.}
Algorithm~\ref{alg:cehs-hm} gives the complete implementation of the
held-out geometry-selection procedure used in the experiments.
\begin{algorithm}[H]
	\caption{CEHS-HM geometry selection}
	\label{alg:cehs-hm}
	\begin{algorithmic}[1]
		
		\Require Data $x_1,\ldots,x_n$; geometry grid $\Lambda$;
		fixed $K$; train-validation splits $\{\mathcal T_\ell,\mathcal V_\ell\}_{\ell=1}^L$.
		\Ensure Selected geometry $\widehat\lambda(K)$.
		
		\For{$\lambda\in\Lambda$}
		\For{$\ell=1,\ldots,L$}
		
		\State Fit energy $K$-medoids on $\mathcal T_\ell$
		under $E_\lambda$, giving medoids $m_1,\ldots,m_K$.
		
		\State Compute the training scale
		\[
		\tau_{\lambda,\ell}
		=
		\operatorname{median}
		\left\{
		E_\lambda(x_i,x_j):
		i<j,\;
		i,j\in\mathcal T_\ell,\;
		E_\lambda(x_i,x_j)>0
		\right\}.
		\]
		
		\State Assign each $x_i\in\mathcal V_\ell$ to its nearest
		training medoid.
		
		\If{$\min_c n_c<2$}
		\State $\widehat C_{\lambda,K,\ell}\gets0$.
		\Else
		
		\For{$c=1,\ldots,K$}
		\State
		$
		\displaystyle
		\widehat H_c
		\gets
		\frac{1}{n_c}
		\sum_{i\in C_c}
		\exp\!\left\{
		-\frac{E_\lambda(x_i,m_c)}
		{\tau_{\lambda,\ell}}
		\right\}.
		$
		\EndFor
		
		\State
		$
		\displaystyle
		\widehat H_{.25}
		\gets
		Q_{.25}(\widehat H_1,\ldots,\widehat H_K).
		$
		
		\For{$1\le c<d\le K$}
		\State
		$
		\displaystyle
		R_{cd}
		\gets
		1-
		\exp\!\left\{
		-\frac{E_\lambda(m_c,m_d)}
		{\tau_{\lambda,\ell}}
		\right\}.
		$
		\EndFor
		
		\State Set $M=K(K-1)/2$ and
		\[
		Q_{\rm HM}
		\gets
		\begin{cases}
			\displaystyle
			\frac{M}{\sum_{c<d}R_{cd}^{-1}},
			& \min_{c<d}R_{cd}>0,\\[6pt]
			0, & \text{otherwise}.
		\end{cases}
		\]
		
		\State
		$
		\displaystyle
		\widehat C_{\lambda,K,\ell}
		\gets
		\frac{2\widehat H_{.25}Q_{\rm HM}}
		{\widehat H_{.25}+Q_{\rm HM}}.
		$
		
		\EndIf
		\EndFor
		
		\State
		$
		\displaystyle
		\widehat C_{\lambda,K}
		\gets
		L^{-1}\sum_{\ell=1}^{L}
		\widehat C_{\lambda,K,\ell}.
		$
		
		\EndFor
		
		\State
		$
		\displaystyle
		\widehat\lambda(K)
		\in
		\arg\max_{\lambda\in\Lambda}
		\widehat C_{\lambda,K}.
		$
		
	\end{algorithmic}
\end{algorithm}

\subsection{Proof of Proposition~\ref{prop:cehs}}

For a valid split, write
$
H=\widehat H_{.25},
\quad
Q=Q_{\rm HM}.
$
By construction,
$
0<H\le1,
\quad
0\le Q<1.
$
Hence
$$
0\le \frac{2HQ}{H+Q}\le1.
$$
Indeed, if $Q=0$, the score is zero. If $H,Q>0$, then
$2HQ/(H+Q)$ is the harmonic mean of $H$ and $Q$, and therefore lies
between $\min(H,Q)$ and $\max(H,Q)$, both of which belong to $(0,1]$.

Next, suppose that all energies at a fixed $\lambda$ are multiplied by
a constant $c>0$. Then the training scale also changes from
$\tau_\lambda$ to $c\tau_\lambda$. Consequently,
$$
\exp\left\{
-\frac{c\E_\lambda}{c\tau_\lambda}
\right\}
=
\exp\left\{
-\frac{\E_\lambda}{\tau_\lambda}
\right\},
$$
so all similarity values, and hence $H$, $Q$, and the resulting CEHS
score, remain unchanged. This proves scale invariance.

Label invariance follows because relabeling the clusters only permutes
the collection of within-cluster quantities entering
$\widehat H_{.25}$ and the collection of pairwise medoid separations
entering $Q_{\rm HM}$. Both the empirical quartile and the harmonic
mean are invariant under such permutations.

For $H,Q>0$,
$$
\frac{\partial}{\partial H}
\frac{2HQ}{H+Q}
=
\frac{2Q^2}{(H+Q)^2}
>0,
$$
and
$$
\frac{\partial}{\partial Q}
\frac{2HQ}{H+Q}
=
\frac{2H^2}{(H+Q)^2}
>0.
$$
Thus the CEHS score is strictly increasing in each component whenever
both components are positive.

Finally, let $R_1,\ldots,R_M>0$. Their harmonic mean is
$$
Q_{\rm HM}
=
\frac{M}
{\displaystyle\sum_{j=1}^M R_j^{-1}}.
$$
Since
$$
\min_j R_j
\le
Q_{\rm HM}
\le
\frac{1}{M}\sum_{j=1}^M R_j,
$$
the harmonic separation cannot fall below the smallest positive
pairwise separation and cannot exceed the corresponding arithmetic
mean. If any $R_j=0$, our convention sets $Q_{\rm HM}=0$, which is
consistent with complete separation failure for at least one medoid
pair.

\section{Final geometry-selection validation studies}
\label{app:cehs_details}
\subsection{Controlled generating deformation}
\label{app:new-controlled-selection}
This study uses the same three-cluster curved-basin generator with $n=300$ ($100$ observations per cluster) and 100 independent replicates at each $\lambda_0\in\{0,.15,.35,.45,.525\}$. The candidate grid is fixed before evaluation. Table~\ref{tab:supp-controlled-selection} reports exact matching only as a diagnostic; it is not used to define the statistical target.

\begin{table}[h]
\centering
\caption{Controlled CEHS selection behavior. Exact $\lambda$ matching is diagnostic, not an estimation claim.}
\label{tab:supp-controlled-selection}
\small
\scalebox{1.17}{\begin{tabular}{rccccc}
\toprule
$\lambda_0$ & $P(\widehat\lambda=\lambda_0)$ & $P(\widehat K=3)$ & MAE$(\widehat\lambda)$ & $E\widehat\lambda$ & $P$(within one grid step)\\
\midrule
0.000 & .06 & 1.00 & .4238 & .4238 & .08\\
0.150 & .03 & 1.00 & .2955 & .4185 & .14\\
0.350 & .02 & 1.00 & .1510 & .4820 & .41\\
0.450 & .03 & 1.00 & .0795 & .4905 & .95\\
0.525 & .82 & 1.00 & .0183 & .5068 & .95\\
\bottomrule
\end{tabular}}
\end{table}

\begin{figure}[h]
\centering
\includegraphics[width=.92\linewidth]{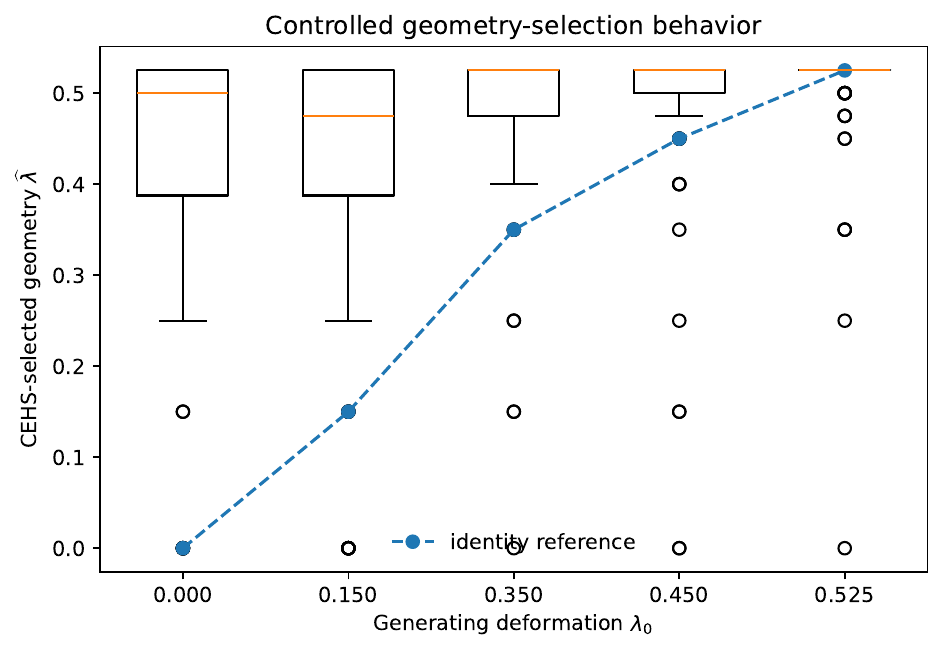}
\caption{Selected CEHS geometry by generating deformation in the 500-dataset controlled study. The dashed identity reference is descriptive only; CEHS optimizes held-out clustering score rather than estimating $\lambda_0$.}
\label{fig:supp-controlled-selection}
\end{figure}

When $\lambda_0=0$, the false-positive nonzero-warping selection rate is $.94$ (Wilson 95\% interval $[.875,.972]$). The mean fixed-$K=3$ CEHS score increases from approximately $.7305$ at candidate $\lambda=0$ to $.7382$ at $\lambda=.525$, showing that the discrepancy is not merely an exact-tie artifact. Figure~\ref{fig:supp-controlled-selection} visualizes the selected-geometry distribution. These results motivate the tuning-parameter interpretation used throughout the main paper.

\subsection{Neutral product-von-Mises misspecification}
\label{app:neutral-study}
The neutral DGP is an equal-weight three-component product-von-Mises mixture on $\T^3$ with $\kappa=2$ and conditional independence across coordinates. Component centers are $(.15,6.05,.25)$, $(2.20,2.00,1.90)$, and $(4.35,4.15,4.50)$ before a common random phase shift that is redrawn each replicate. This phase shift avoids accidental alignment with the fixed angular origin of the CTE family while leaving the flat product-torus problem unchanged. There are $100$ observations per component and $100$ replicates. No CTE metric or $\lambda$ enters the sampling mechanism.

With $K=3$ fixed, $P(\widehat\lambda=0)=.56$ (Wilson 95\% interval $[.462,.653]$) and $P(\widehat\lambda\le.15)=.74$ (interval $[.646,.816]$); the mean and median selected values are $.11575$ and $0$. Mean ARI is $.8138$ at the flat endpoint and $.8072$ under selected geometry. The paired difference is $-.006595$ (95\% $t$ interval $[-.011784,-.001405]$); $47\%$ of replicates, it ties the flat ARI exactly and $84\%$ remains within $.02$ of it. The label-based oracle gain available from the candidate geometry family is only $.01146$ ARI on average.

The joint extension behaves differently; it recovers $K=3$ in only $.06$ one replicate (Wilson interval $[.0278,.1248]$), with the mean selected $K=2.06$. Its paired ARI difference from flat $K=3$ is $-.36583$ (95\% interval $[-.38481,-.34685]$). At $\lambda=0$, mean CEHS is $.77899$ for $K=2$ versus $.76585$ for $K=3$, and the coarser partition remains favored across the geometry grid. This is why joint $K$ selection is treated as a secondary
finite-grid, data-adaptive CEHS-HM extension rather than as an
estimator of a generative component count.

\begin{figure}[h]
\centering
\includegraphics[width=.92\linewidth]{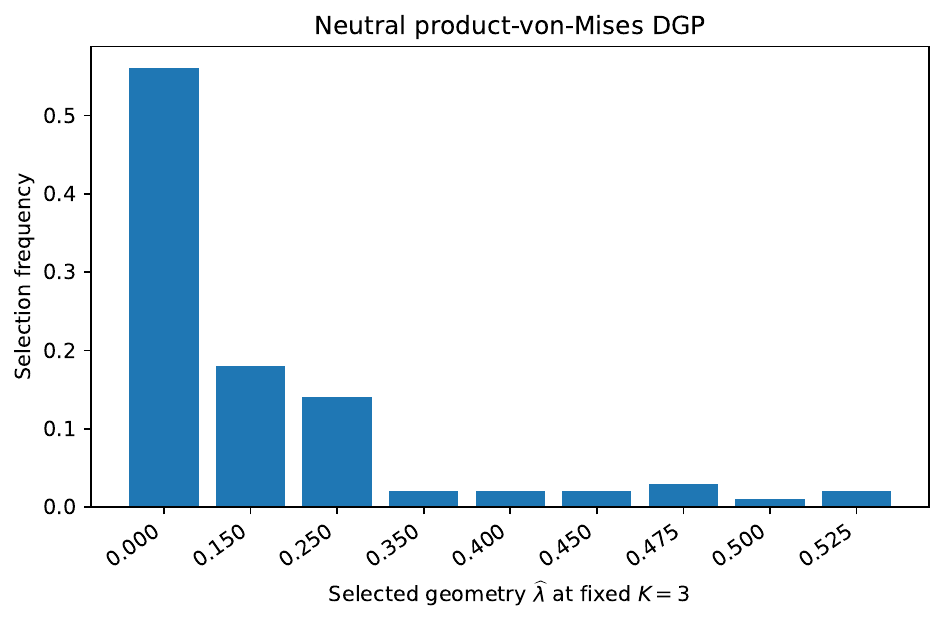}
\caption{Fixed-$K=3$ CEHS geometry-selection frequencies for the neutral product-von-Mises DGP.}
\label{fig:supp-neutral-selection}
\end{figure}

\subsection{Held-out baseline criteria and detailed comparison}
\label{app:baseline-criteria}

The  confirmatory study compares CEHS-HM with two
label-free cluster-validity criteria: held-out silhouette
\citep{rousseeuw1987silhouettes} and a pairwise
energy-dispersion pseudo-$F$ criterion. The latter follows the
general idea of a between-to-within dispersion ratio
\citep{calinski1974dendrite}, expressed here directly through
pairwise CTE energies. Related distance-based pseudo-$F$
constructions can be formulated from a symmetric dissimilarity
matrix \citep{anderson2001new}. No $F$-distribution interpretation
is used.

\paragraph{Common held-out protocol.}
For every candidate $(\lambda,K)$ and every train-validation split,
the three criteria use exactly the same CTE distance matrix and the
same energy $K$-medoids fit on the training observations.
Validation observations are assigned to their nearest fitted
training medoids. Thus the three methods differ only in the score
applied to the resulting held-out partition.

For invalid held-out evaluations, the criterion-specific minimum
score is used. If any fitted training cluster receives fewer than
two validation observations, CEHS-HM and pseudo-$F$ receive score
zero, while silhouette receives $-1$. No invalid held-out
evaluation is discarded when candidate scores are averaged.

\paragraph{Held-out silhouette.}
Let $a_i$ denote the mean CTE dissimilarity from validation
observation $i$ to other observations in its assigned cluster, and
let $b_i$ denote the smallest mean dissimilarity from $i$ to any
other validation cluster. The silhouette value is
$$
s_i
=
\frac{b_i-a_i}
{\max(a_i,b_i)},
$$
and the held-out silhouette criterion is the mean of $s_i$ over the
validation observations. Cluster memberships are determined only by
the training medoids; the validation data are not reclustered.

\paragraph{Pairwise energy-dispersion pseudo-$F$.}
For a validation sample of size $n_v$, define
$$
E_{ij}
=
D_{E,\lambda}(x_i,x_j)^2.
$$
For the validation partition $C_1,\ldots,C_K$, let
$$
T
=
\frac{1}{2n_v}
\sum_{i,j}E_{ij},
$$
and
$
W
=
\sum_{c=1}^{K}
\frac{1}{2n_c}
\sum_{i,j\in C_c}E_{ij}.
$
Writing
$
B=T-W,
$
the pseudo-$F$ criterion is
$$
F^*
=
\frac{B/(K-1)}
{W/(n_v-K)}.
$$
This quantity is used only as a cluster-validity score; it is not
assumed to follow an $F$ distribution.

\paragraph{Candidate selection.}
For each criterion $S$, the scores are averaged over all
prespecified held-out evaluations,
$$
\overline S_{\lambda,K}
=
\frac{1}{L}
\sum_{\ell=1}^{L}
S_{\lambda,K,\ell},
$$
and the selected candidate is
$$
(\widehat\lambda,\widehat K)
\in
\arg\max_{(\lambda,K)}
\overline S_{\lambda,K}.
$$
The same held-out splits and candidate grid are used for all three
criteria.

\paragraph{Scenario-level results.}
The independent confirmation contains $250$ fresh replicates in
each of four prespecified scenarios. CEHS-HM shows its largest
advantage in the unbalanced three-cluster setting, whereas
pseudo-$F$ or silhouette can have slightly smaller ARI regret in
some balanced settings. Thus the aggregate comparison in
Table~\ref{tab:cehs_confirmatory_main} should not be interpreted as
uniform superiority of CEHS-HM across every data-generating
configuration. Table~\ref{tab:cehs_confirmatory_scenarios} gives the
complete scenario-level results.
\begin{table*}[h!]
	\centering
	\caption{Scenario-level confirmatory model-selection results.
		ARI regret is the difference between the independent-test oracle
		ARI and the ARI of the selected candidate. Severe regret denotes
		ARI regret at least $0.10$.}
	\label{tab:cehs_confirmatory_scenarios}
	\small
 \scalebox{1.07}{	\begin{tabular}{llccc}
		\toprule
		Scenario & Criterion &
		$P(\widehat K=K_0)$ &
		Mean ARI regret &
		Severe regret \\
		\midrule
		Moderate $K=3$
		& CEHS-HM   & 0.992 & 0.0201 & 0.012 \\
		& Pseudo-$F$ & 1.000 & 0.0159 & 0.004 \\
		& Silhouette & 1.000 & 0.0149 & 0.000 \\[2pt]
		
		Hard $K=3$
		& CEHS-HM   & 0.984 & 0.0288 & 0.048 \\
		& Pseudo-$F$ & 0.988 & 0.0291 & 0.052 \\
		& Silhouette & 0.928 & 0.0444 & 0.104 \\[2pt]
		
		Four clusters
		& CEHS-HM   & 0.988 & 0.0224 & 0.032 \\
		& Pseudo-$F$ & 0.988 & 0.0196 & 0.028 \\
		& Silhouette & 0.980 & 0.0220 & 0.028 \\[2pt]
		
		Unbalanced $K=3$
		& CEHS-HM   & 0.988 & 0.0338 & 0.056 \\
		& Pseudo-$F$ & 0.812 & 0.0656 & 0.224 \\
		& Silhouette & 0.780 & 0.0698 & 0.240 \\
		\bottomrule
	\end{tabular}}
\end{table*}
The scenario-level results clarify the source of the aggregate
difference. In the moderately balanced setting, pseudo-$F$ and
silhouette have slightly smaller mean ARI regret than CEHS-HM.
The three criteria perform similarly in the four-cluster setting.
In contrast, CEHS-HM is more robust in the hard and especially
the unbalanced three-cluster setting. In the latter, CEHS-HM
selects the generating resolution in $98.8\%$ of replicates,
compared with $81.2\%$ for pseudo-$F$ and $78.0\%$ for silhouette.
Its severe-regret rate is also substantially smaller in this
setting.

The prespecified aggregate CEHS-HM minus pseudo-$F$
mean-regret contrast is $-0.00622$, with a $95\%$ stratified
paired-bootstrap interval $[-0.00965,-0.00276]$.
The improvement is driven primarily by the difficult and unbalanced
settings rather than by uniform dominance across all scenarios.

    \subsection{Two-dimensional benchmark details}
\label{app:t2_details}

The benchmark uses three clusters and $n=300$ observations per
replicate. The curved-basin generator produces approximately compact
clusters in the curved geometry, whereas the curved-band generator
produces nonlinear one-dimensional structures.

\begin{figure*}[t]
    \centering
    \includegraphics[width=0.9\textwidth]
    {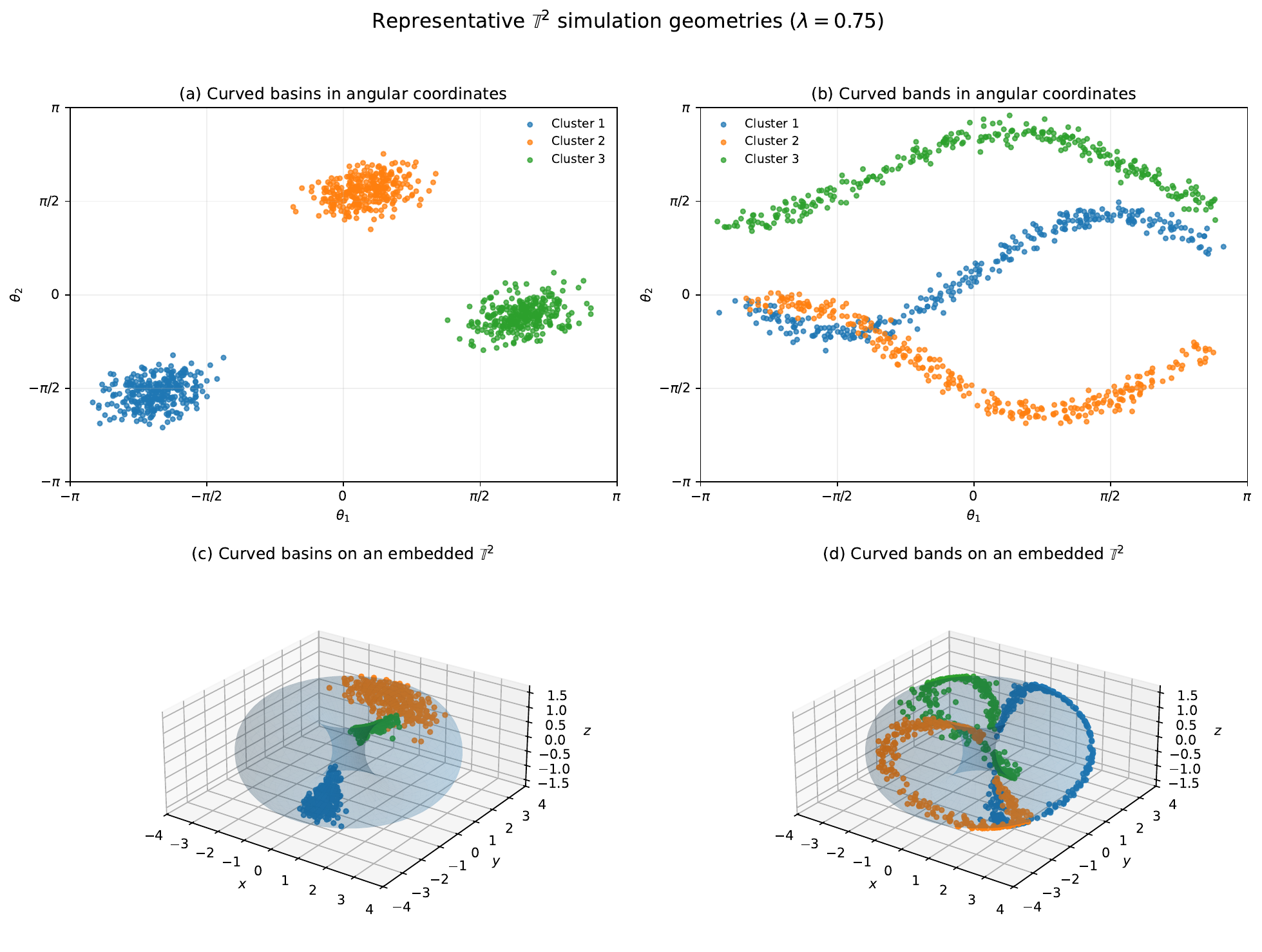}
    \caption{
        Representative artificial data used in the
        $\mathbb{T}^2$ clustering experiments.
        Top: angular-coordinate representations of the compact curved-basin
        and nonlinear curved-band designs.
        Bottom: the same observations displayed on an embedded torus to
        emphasize their periodic topology.
        The three-dimensional embedding is used only for visualization;
        clustering is performed using the intrinsic CTE geometry with
        $\lambda=0.75$.
    }
    \label{fig:supp_t2_simulation_full}
\end{figure*}

Representative realizations of the two benchmark designs are shown in
Figure~\ref{fig:supp_t2_simulation_full}. The upper panels display the
observations in angular coordinates, while the lower panels show the same
observations on an embedded torus. The embedding is used only for
visualization; all clustering calculations use the CTE dissimilarity.

    \begin{table}[h]
        \centering
        \caption{$\T^2$ benchmark: mean ARI (SD) over 30 paired replicates. MINT-AGE denotes the dimension-general published-algorithm reimplementation described below.}
        \label{tab:supp_t2_full}
        \small
        	\scalebox{1.2}{\begin{tabular}{lcc}
            \toprule
            Method & Curved basins & Curved bands\\
            \midrule
            CTE Fr\'echet $K$-means & $0.769\ (0.055)$ & $0.374\ (0.086)$\\
            CTE energy-medoid & $0.767\ (0.053)$ & $0.349\ (0.085)$\\
            CTE spectral & $0.714\ (0.072)$ & $0.335\ (0.131)$\\
            CTE hierarchical & $0.672\ (0.149)$ & $0.419\ (0.137)$\\
            CTE HDBSCAN & $0.605\ (0.196)$ & $0.578\ (0.203)$\\
            CTE DBSCAN & $0.256\ (0.280)$ & $0.494\ (0.074)$\\
            PSM-DBSCAN & $0.544\ (0.196)$ & $0.520\ (0.115)$\\
            Flat DBSCAN & $0.229\ (0.294)$ & $0.485\ (0.061)$\\
            MINT-AGE reimpl. & $0.414\ (0.209)$ & $0.203\ (0.074)$\\
            \bottomrule
        \end{tabular}}
    \end{table}

\begin{figure}[htbp] % 'htbp' prevents the "Float too large / changed to ht" warnings
	\centering
	\begin{minipage}{0.48\textwidth}
		\centering
		\includegraphics[width=\linewidth]{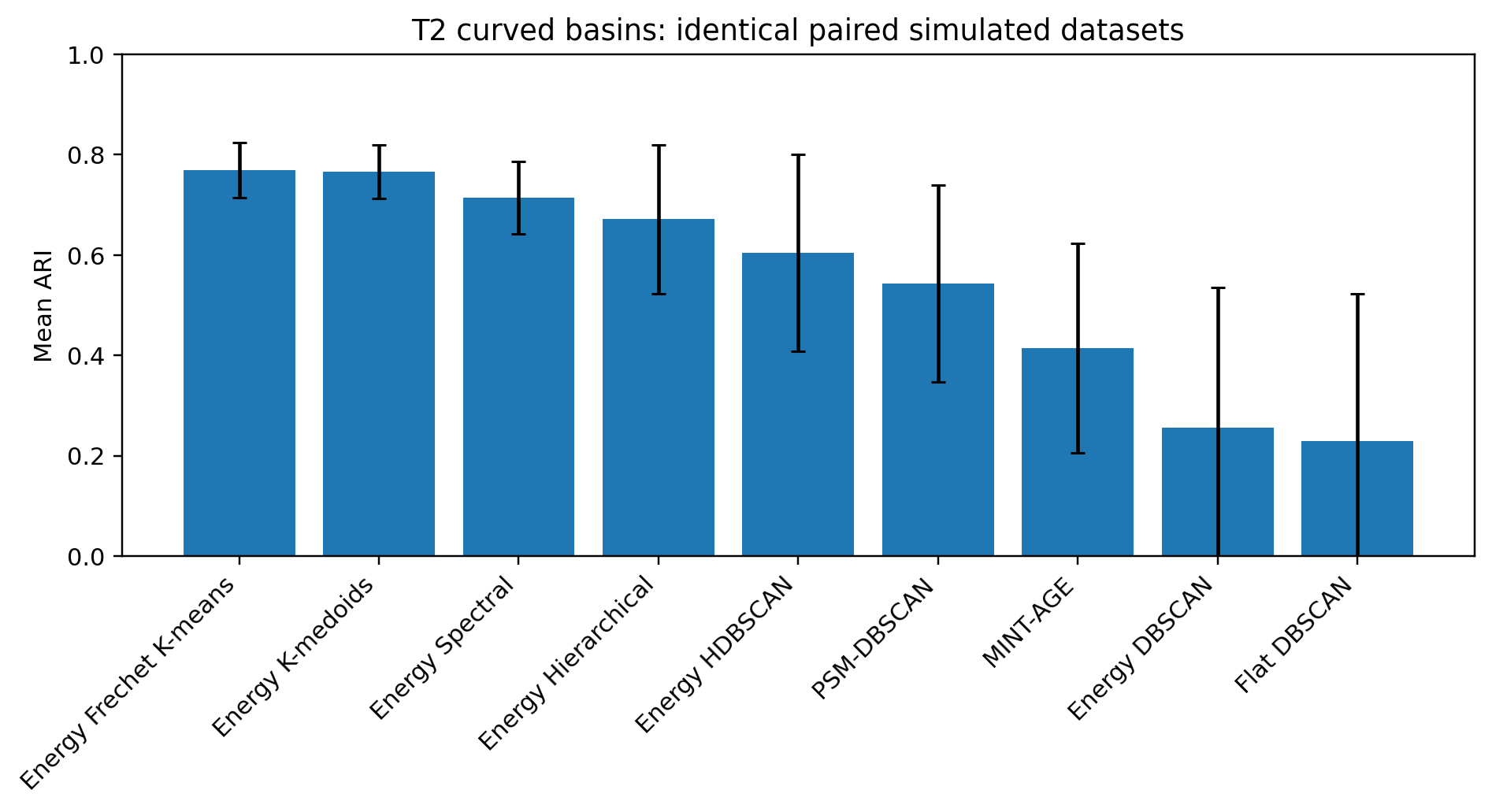}
		\caption{Curved basins.}
	\end{minipage}\hfill
	\begin{minipage}{0.48\textwidth}
		\centering
		\includegraphics[width=\linewidth]{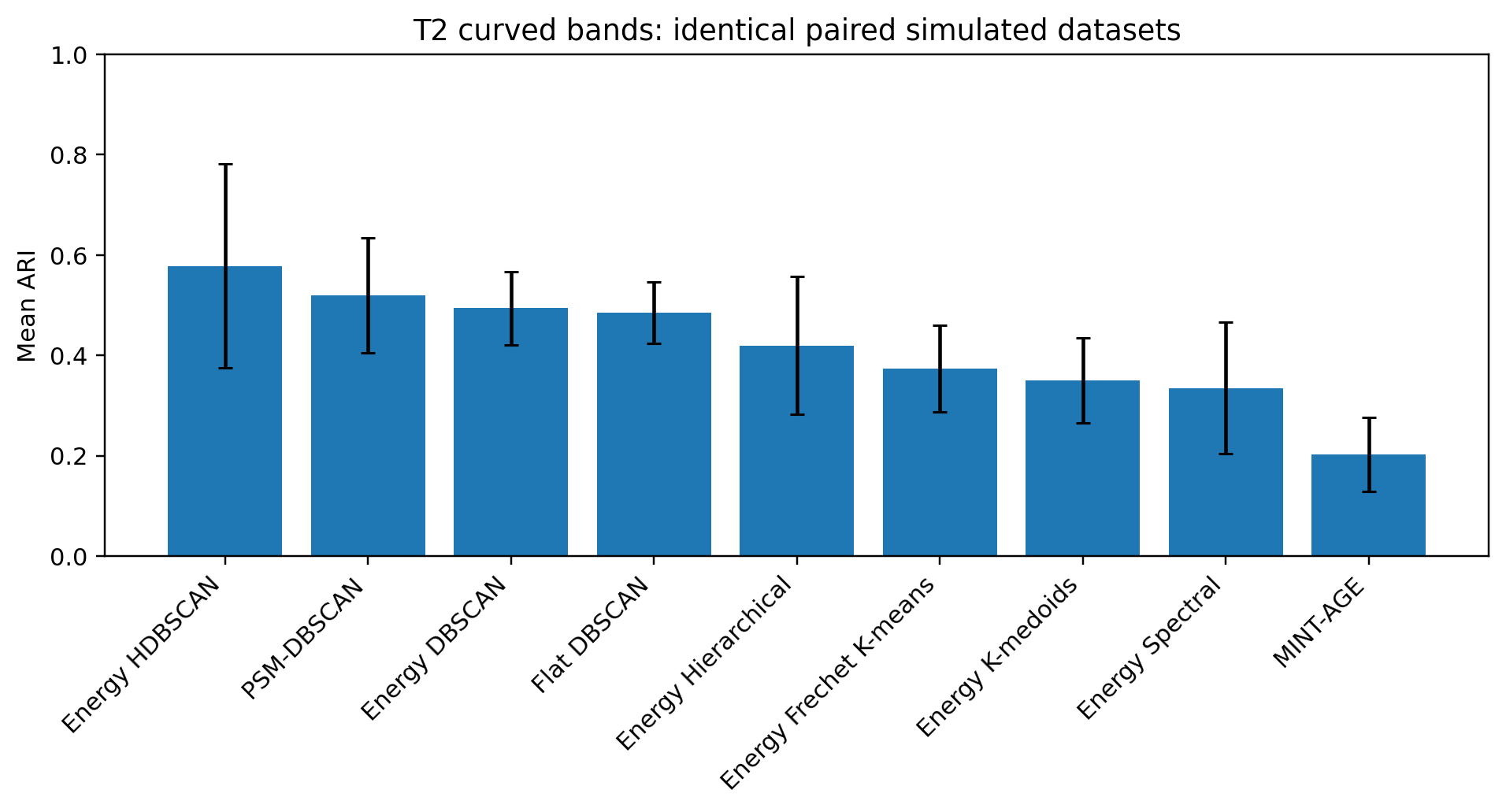}
		\caption{Curved bands.}
	\end{minipage}
	\caption{Final $\T^2$ benchmark figures from the 30-replicate output.}
	\label{fig:supp_t2} % <-- Added missing closing brace here
\end{figure}

%    \begin{table}[h]
%        \centering
%        \caption{Paired differences versus PSM-DBSCAN over 30 replicates.}
%        \scalebox{.71}{    \begin{tabular}{llcc}
%                
%                \toprule
%                Scenario & CTE method & Mean difference (95\% CI) & Win rate\\
%                \midrule
%                Basins & Fr\'echet $K$-means & $0.177\ [0.115,0.240]$ & 0.933\\
%                Basins & Energy-medoid & $0.173\ [0.111,0.236]$ & 0.933\\
%                Basins & Spectral & $0.123\ [0.066,0.181]$ & 0.767\\
%                Bands & HDBSCAN & $0.068\ [0.008,0.129]$ & 0.567\\
%                Bands & DBSCAN & $0.006\ [-0.014,0.025]$ & 0.333\\
%                Bands & Energy-medoid & $-0.085\ [-0.118,-0.052]$ & 0.267\\
%                \bottomrule
%        \end{tabular}}
%    \end{table}
%    
    \begin{table}[t]
        \centering
        \caption{Selected paired ARI differences relative to PSM-DBSCAN over
            $30$ replicates. Positive differences favor the CTE-based method.}
            \label{tab:supp_t2_paired_psm}
    \scalebox{1.17804}    {\begin{tabular}{llcc}
            \toprule
            Scenario & CTE method & Mean difference (95\% CI) & Win rate \\
            \midrule
            Basins & Fr\'echet K-means
            & 0.225 $[0.161,\,0.289]$ & 0.967 \\
            Basins & Energy-medoid
            & 0.223 $[0.158,\,0.287]$ & 0.967 \\
            Basins & Spectral
            & 0.170 $[0.102,\,0.238]$ & 0.800 \\
            Bands & HDBSCAN
            & 0.058 $[-0.017,\,0.134]$ & 0.533 \\
            Bands & DBSCAN
            & -0.026 $[-0.066,\,0.014]$ & 0.233 \\
            Bands & Energy-medoid
            & -0.170 $[-0.221,\,-0.120]$ & 0.067 \\
            \bottomrule
        \end{tabular}}
    \end{table}

    Table~\ref{tab:supp_t2_paired_psm} reports the selected paired contrasts with PSM-DBSCAN. In curved basins, the Fr\'echet, energy-medoid, and spectral CTE procedures all have positive paired intervals. In curved bands, CTE-HDBSCAN has the largest mean ARI in Table~\ref{tab:supp_t2_full}, but its paired difference relative to PSM-DBSCAN is $0.058$ with interval $[-0.017,0.134]$, so a clear paired advantage is not established; energy-medoid clustering is substantially worse than PSM-DBSCAN for this morphology. For visual confirmation, see Figure~\ref{fig:supp_t2} for the mean ARI across the different clustering methods evaluated in this study.
    
    The PSM implementation in the supplied benchmark is an independent implementation following the published principal-submanifold update structure \cite[see][]{wuyao2026}. The MINT-AGE comparator is a dimension-general reimplementation of the published AGE-TOSS-PNS-mode-hunting architecture \citep{mardia2022pca}. MINT-AGE estimates its own number of clusters, whereas the known-$K$ center/medoid simulations use $K=3$; accordingly these are supporting complete-procedure comparisons rather than fixed-$K$ geometry ablations. Its mean selected cluster count is $2.83$ in curved basins and $15.63$ in curved bands, explaining the strong over-segmentation observed in the nonlinear-band design.
    
    The official ClusTorus 0.2.2 package was applied separately. Its known-$K$ analysis searched mixture component counts $J=1,\ldots,15$ and retained candidate outputs with three final clusters. The package contains 100 ClusTorus replicates; paired comparisons use the first 30 matched replicates. For curved basins, CTE Fr\'echet $K$-means exceeds the ClusTorus Ae and Ao assignments by mean ARI differences $0.490$ (95\% CI $[0.395,0.584]$) and $0.541$ ($[0.466,0.616]$), respectively. For curved bands, CTE-HDBSCAN exceeds the corresponding Ae and Ao assignments by $0.113$ ($[0.036,0.189]$) and $0.123$ ($[0.046,0.200]$), respectively. Across all 100 ClusTorus replicates, mean ARIs are $0.2615$ (Ae) and $0.2177$ (Ao) in curved basins, and $0.4834$ and $0.4725$ in curved bands.

    %     \begin{figure*}[h]
    %     \centering
    %     \includegraphics[width=0.9\textwidth]{figures/T2_curved_basins_bands_angular_and_torus.pdf}
    %     \caption{
    %         Representative artificial data used in the
    %         $\mathbb{T}^2$ clustering experiments.
    %         Top: angular-coordinate representations of the compact curved-basin
    %         and nonlinear curved-band designs.
    %         Bottom: the same observations displayed on an embedded torus to
    %         emphasize their periodic topology.
    %         The three-dimensional embedding is used only for visualization;
    %         clustering is performed using the intrinsic CTE geometry with
    %         $\lambda=0.75$.
    %     }
    %     \label{fig:supp_t2_simulation_full}
    % \end{figure*}
    
    \section{Warping-study details}
    
\begin{figure}[h]
        \centering
        \includegraphics[width=0.8\textwidth]{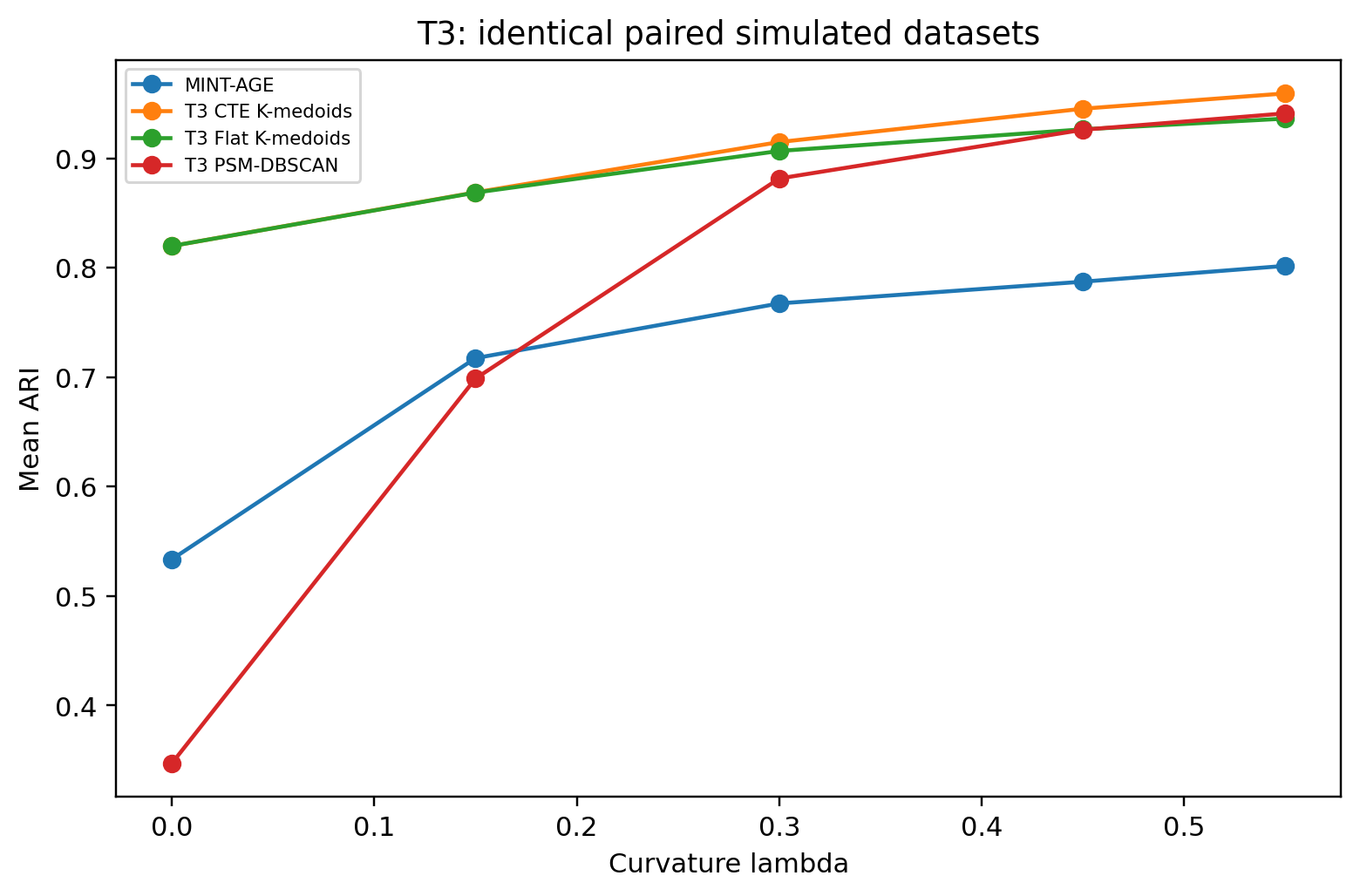}
        \caption{Absolute mean ARI as a function of $\lambda$ for CTE, flat $K$-medoids, PSM-DBSCAN, and the dimension-general MINT-AGE reimplementation.}
        \label{fig:supp_t3_full}
    \end{figure}

    The 100-replicate $\T^3$ study uses the same latent random seed for all $\lambda$ values within a replicate. This pairing is essential; observed changes in $\ARI_{\mathrm{CTE}}-\ARI_{\mathrm{flat}}$ are not due to independently regenerated noise. Table~\ref{tab:supp_t3_full} reports the absolute mean ARIs, and Figure~\ref{fig:supp_t3_full} displays the same comparison across the five warping levels.

        \begin{table*}[h]
        \centering
        \caption{$\T^3$ mean ARI (SD) by warping level.}
        \label{tab:supp_t3_full}
        \renewcommand{\arraystretch}{1.5} % Increases row height by 30%
        \scalebox{1}{
            \begin{tabular}{ccccc}
                \toprule
                $\lambda$ & CTE energy-medoid & Flat energy-medoid & PSM-DBSCAN & MINT-AGE reimpl.\\
                \midrule
                0.00 & $0.8199\ (0.0435)$ & $0.8199\ (0.0435)$ & $0.3470\ (0.3850)$ & $0.5335\ (0.2712)$\\
                0.15 & $0.8690\ (0.0356)$ & $0.8687\ (0.0373)$ & $0.6988\ (0.3238)$ & $0.7174\ (0.1813)$\\
                0.30 & $0.9149\ (0.0280)$ & $0.9068\ (0.0293)$ & $0.8816\ (0.1032)$ & $0.7675\ (0.1649)$\\
                0.45 & $0.9454\ (0.0236)$ & $0.9266\ (0.0231)$ & $0.9262\ (0.0294)$ & $0.7873\ (0.1442)$\\
                0.55 & $0.9594\ (0.0198)$ & $0.9364\ (0.0220)$ & $0.9411\ (0.0448)$ & $0.8018\ (0.1372)$\\
                \bottomrule
            \end{tabular}
        }
    \end{table*}
    
%    \begin{table}[h]
%        \centering
%        \caption{$\T^3$ mean ARI (SD) by warping level.}
%        \label{tab:supp_t3_full}
%        \scalebox{.6}{    \begin{tabular}{ccccc}
%                \toprule
%                $\lambda$ & CTE energy-medoid & Flat energy-medoid & PSM-DBSCAN & MINT-AGE reimpl.\\
%                \midrule
%                0.00 & $0.8199\ (0.0435)$ & $0.8199\ (0.0435)$ & $0.3470\ (0.3850)$ & $0.5335\ (0.2712)$\\
%                0.15 & $0.8690\ (0.0356)$ & $0.8687\ (0.0373)$ & $0.6988\ (0.3238)$ & $0.7174\ (0.1813)$\\
%                0.30 & $0.9149\ (0.0280)$ & $0.9068\ (0.0293)$ & $0.8816\ (0.1032)$ & $0.7675\ (0.1649)$\\
%                0.45 & $0.9454\ (0.0236)$ & $0.9266\ (0.0231)$ & $0.9262\ (0.0294)$ & $0.7873\ (0.1442)$\\
%                0.55 & $0.9594\ (0.0198)$ & $0.9364\ (0.0220)$ & $0.9411\ (0.0448)$ & $0.8018\ (0.1372)$\\
%                \bottomrule
%        \end{tabular}}
%    \end{table}

    For the MINT-AGE reimplementation, the mean selected cluster counts are $2.70$, $2.80$, $2.96$, $3.40$, and $3.47$ as $\lambda$ increases from $0$ to $0.55$. This automatic cluster-number selection is not used in the controlled CTE-versus-flat mechanism contrast.
    
    \section{Global geodesic robustness and \texorpdfstring{$\mathbb T^N$}{T\textasciicircum N} computational scalability}
    \label{app:geodesic}

\begin{figure}[!htbp]
	\centering
	\begin{minipage}{0.48\textwidth}
		\centering
		\includegraphics[width=\linewidth]{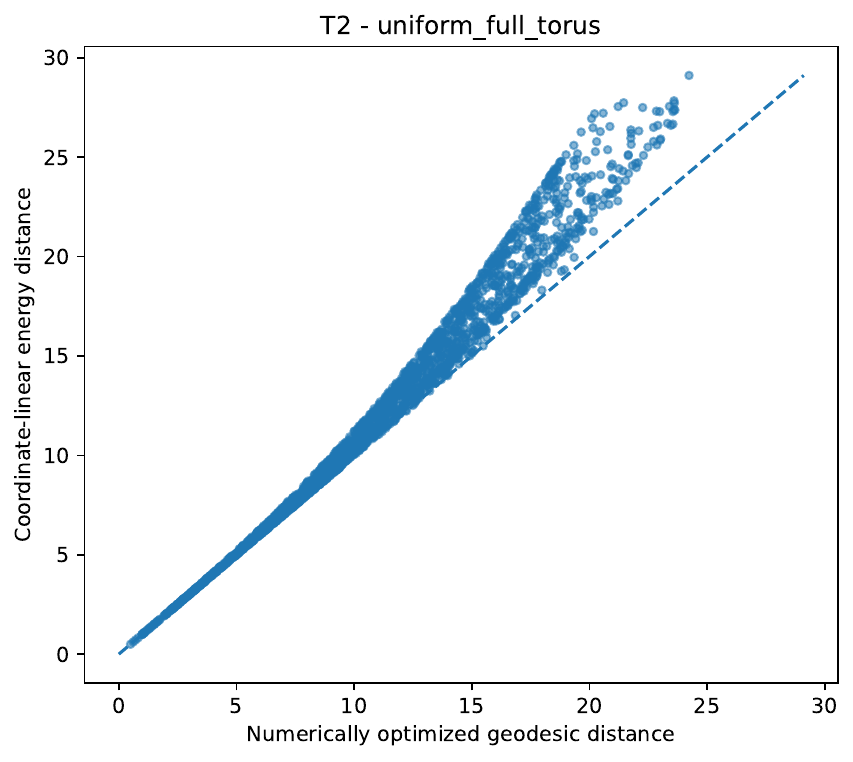}
		\caption{$\T^2$, uniform.}
	\end{minipage}\hfill
	\begin{minipage}{0.48\textwidth}
		\centering
		\includegraphics[width=\linewidth]{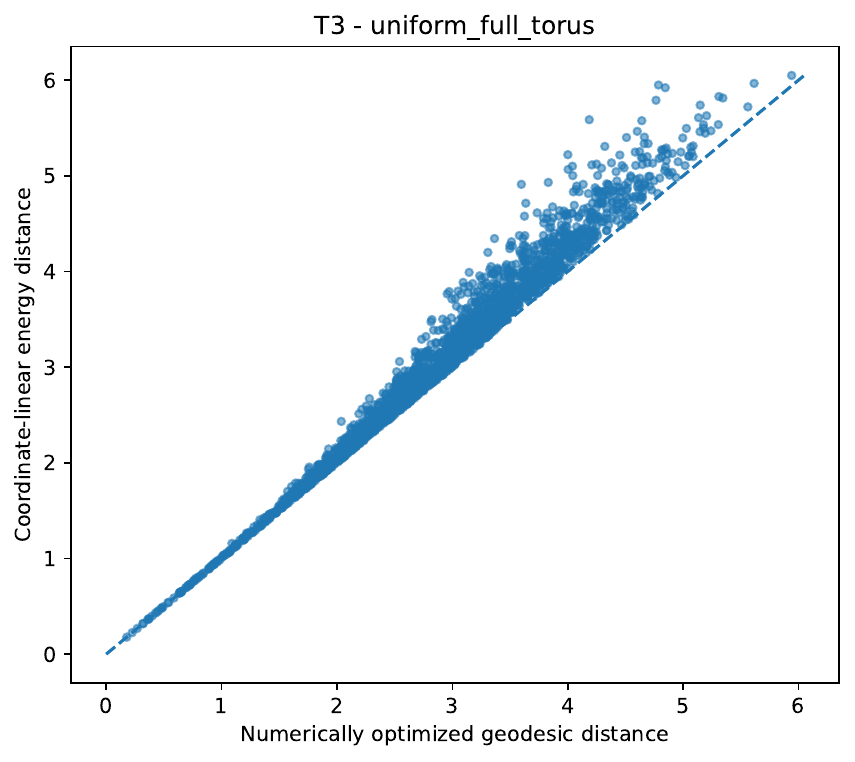}
		\caption{$\T^3$, uniform.}
	\end{minipage}
	\caption{Global full-torus validation. The surrogate is an upper-bound dissimilarity, not an everywhere-negligible approximation.}
	\label{fig:geodesic_uniform}
\end{figure}

\begin{figure}[!htbp]
	\centering
	\begin{minipage}{0.48\textwidth}
		\centering
		\includegraphics[width=\linewidth]{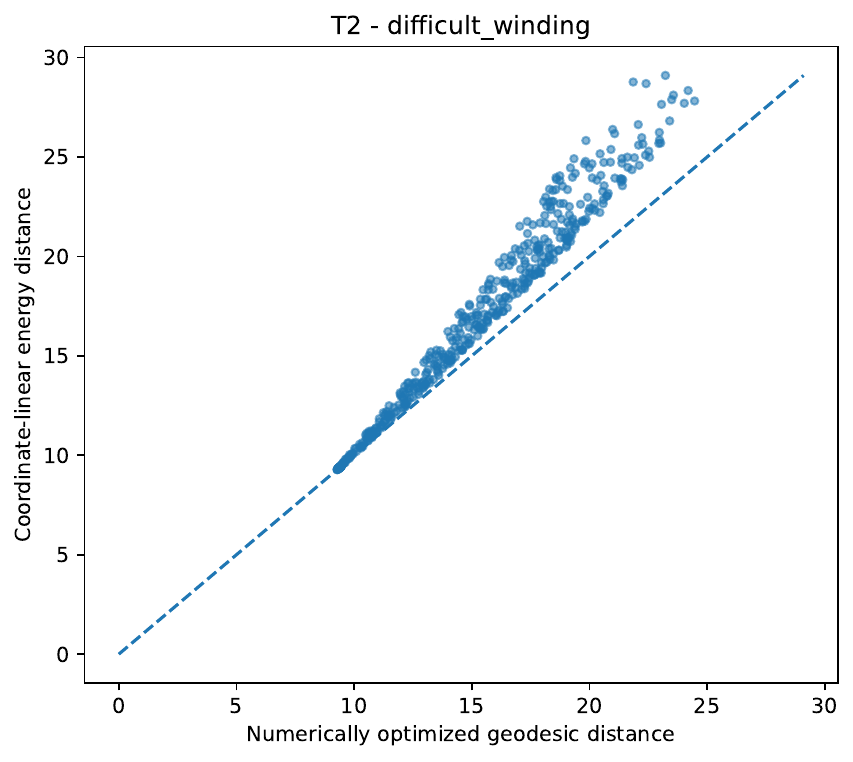}
		\caption{$\T^2$, difficult winding.}
	\end{minipage}\hfill
	\begin{minipage}{0.48\textwidth}
		\centering
		\includegraphics[width=\linewidth]{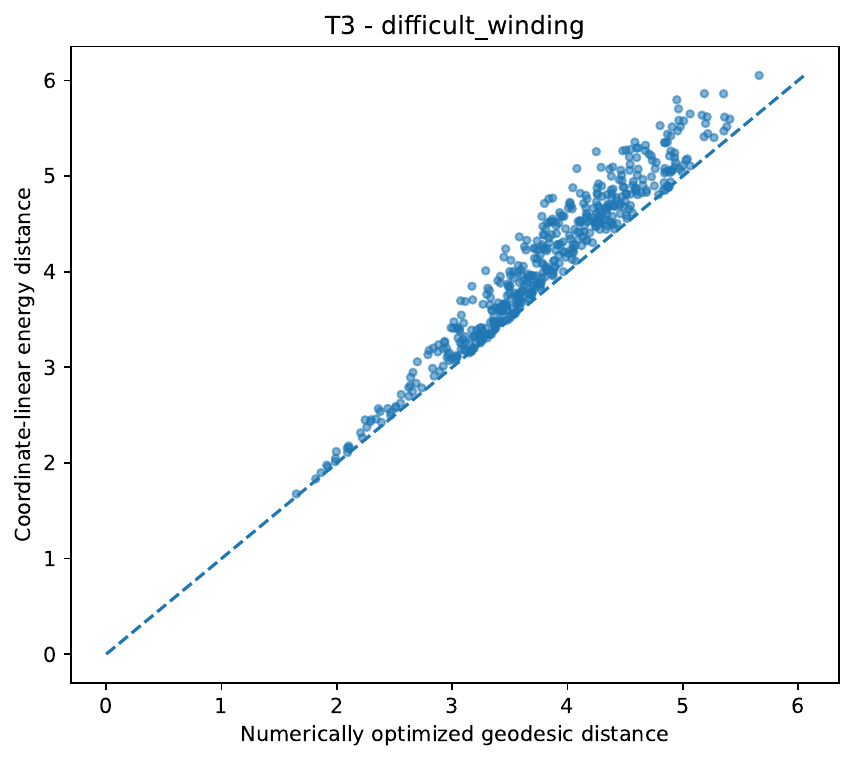}
		\caption{$\T^3$, difficult winding.}
	\end{minipage}
	\caption{
		CTE versus numerically optimized geodesic distance for deliberately difficult
		winding pairs on $\T^2$ and $\T^3$. The experiment stresses nontrivial winding
		configurations and assesses how well CTE preserves geodesic behavior in these
		challenging cases.
	}
	\label{fig:geodesic_difficult}
\end{figure}

    Table~\ref{tab:supp_geodesic} summarizes the global and difficult-winding comparisons. Figures~\ref{fig:geodesic_uniform} and~\ref{fig:geodesic_difficult} show the corresponding distance scatterplots. There were zero numerical cases in which $\dg^{\mathrm{num}}$ exceeded $\DE$ beyond tolerance. Candidate winding sets became larger by design in the difficult experiment: the mean numbers were $3.042$ in $\T^2$ and $6.504$ in $\T^3$, compared with $1.792$ and $4.048$ under uniform full-torus sampling.

A dimension study considers
$N\in\{2,3,5,7,10\}$ with $\lambda=.40$, $K=3$, $n=300$,
and 100 replicates at each dimension. The principal numerical
summary is reported in Table~\ref{tab:scaling-main} of the main
paper.

The finite-winding search shows no substantial combinatorial
growth in these compact-cluster experiments: the replicate-level
95th percentile number of candidates per pair is one at every
dimension, while the maximum observed numbers are
$2,2,5,7,$ and $13$ for $N=2,3,5,7,10$, respectively.

\begin{figure}[!htbp]
	\centering
	\begin{minipage}{0.48\textwidth}
		\centering
		\includegraphics[width=\linewidth]{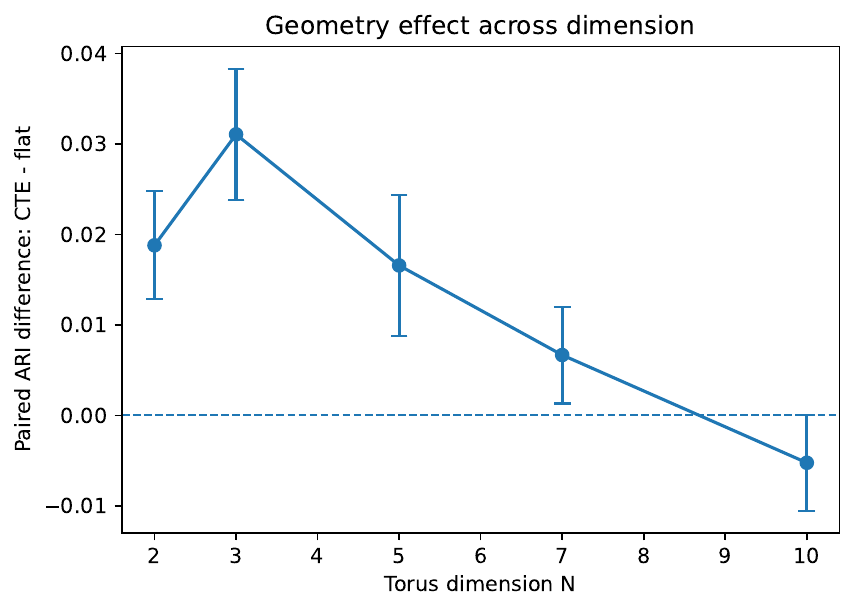}
		\caption{Paired ARI difference.}
	\end{minipage}\hfill
	\begin{minipage}{0.48\textwidth}
		\centering
		\includegraphics[width=\linewidth]{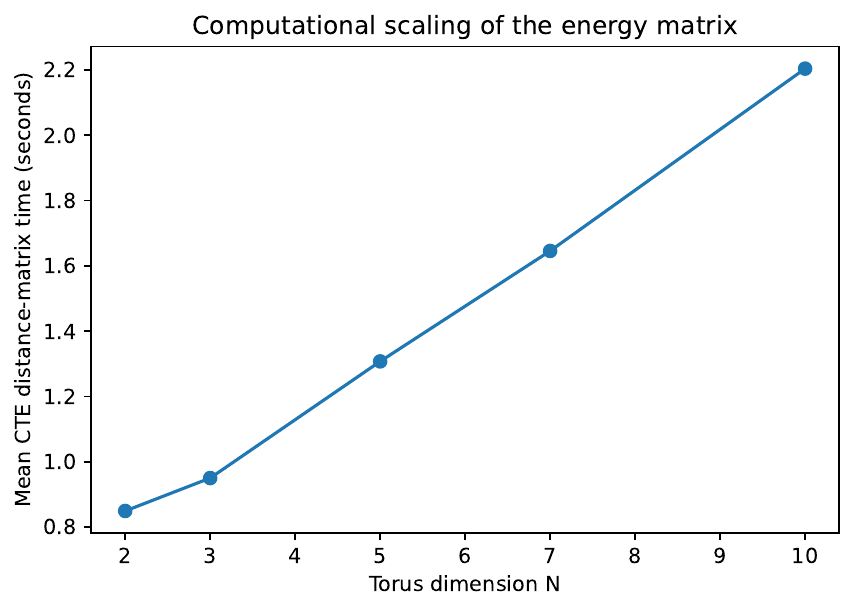}
		\caption{CTE matrix runtime.}
	\end{minipage}
	\caption{
		Dimension study for CTE on $\mathbb{T}^N$.
		Panel (a) shows the paired ARI difference between CTE and flat-torus
		clustering, and panel (b) reports the mean time required to construct the
		CTE distance matrix as the torus dimension increases.
	}
	\label{fig:supp_tn_scalability}
\end{figure}

Figure~\ref{fig:supp_tn_scalability} displays the paired CTE-minus-flat
ARI effect and empirical CTE distance-matrix runtime. At $N=10$,
the paired difference is $-0.0052$ with 95\% CI
$[-0.0106,0.0001]$, so an improvement over the flat geometry is 
not established at that dimension.

    % Float64 CPU refinement of the 50 hardest pairs in each geometry returned raw SciPy success for every case, mesh convergence for every case, preservation of the theoretical upper-bound direction for every case, and GPU-CPU agreement for every case. Median relative GPU-CPU discrepancies were approximately $3.8\times10^{-5}$ in $\T^2$ and $5.3\times10^{-5}$ in $\T^3$. A separate stringent gradient-norm flag was not met in all runs; this does not contradict the optimizer's objective- and mesh-based checks and is reported transparently in the raw files.
    
            \begin{table*}[h]
        \centering
        \caption{Global and difficult-winding comparison of CTE to numerical geodesics. Gap is $(\DE-\dg^{\mathrm{num}})/\dg^{\mathrm{num}}$.}
        \label{tab:supp_geodesic}
        \scalebox{1.04}{\begin{tabular}{llccccc}
                \toprule
                Geom. & Design & $n$ & Spearman & Median gap & 95th pct. & Wind agreement\\
                \midrule
                $\T^2$ & uniform & 2000 & 0.9950 & 0.0548 & 0.2417 & 0.969\\
                $\T^2$ & difficult winding & 500 & 0.9905 & 0.0815 & 0.2391 & 0.842\\
                $\T^3$ & uniform & 2000 & 0.9890 & 0.0449 & 0.1660 & 0.987\\
                $\T^3$ & difficult winding & 500 & 0.9663 & 0.0619 & 0.1729 & 0.904\\
                \bottomrule
        \end{tabular}}
    \end{table*}

%        \begin{table}[h]
%        \centering
%        \caption{Supplementary $\T^N$ study: same-engine CTE versus flat energy-medoid clustering.}
%        \label{tab:supp_tn_scalability}
%        \scalebox{.78}{    \begin{tabular}{ccccc}
%                \toprule
%                $N$ & CTE ARI & Flat ARI & Paired difference (95\% CI) & CTE matrix sec.\\
%                \midrule
%                2 & 0.4988 & 0.4800 & $0.0188\ [0.0128,0.0248]$ & 0.849\\
%                3 & 0.6424 & 0.6113 & $0.0311\ [0.0238,0.0383]$ & 0.950\\
%                5 & 0.7988 & 0.7823 & $0.0166\ [0.0088,0.0244]$ & 1.307\\
%                7 & 0.8877 & 0.8810 & $0.0067\ [0.0013,0.0120]$ & 1.646\\
%                10 & 0.9429 & 0.9481 & $-0.0052\ [-0.0106,0.0001]$ & 2.204\\
%                \bottomrule
%        \end{tabular}}
%    \end{table}
%    

%    As shown in Table~\ref{tab:supp_tn_scalability} and Figure~\ref{fig:supp_tn_scalability}, the paired CTE-minus-flat interval remains positive through $N=7$. At $N=10$ the point estimate is slightly negative, $-0.0052$, but the interval $[-0.0106,0.0001]$ includes zero. Mean CTE distance-matrix time increases from $0.849$ seconds at $N=2$ to $2.204$ seconds at $N=10$ on the reported run; these timings are hardware-dependent.
%

%\newpage
\section{Additional real-data details}
\label{app:realdata}

\subsection{ILE protein torsion data}
\label{app:ile_details}

\paragraph{Data and analysis.}
The ILE data contain $8080$ complete observations on
$(\phi,\psi,\chi_1,\chi_2)\in\mathbb{T}^4$.
The coordinate order is kept fixed throughout because the recursive
CTE geometry assigns a specific geometric role to coordinate
position. Model selection is entirely label-blind; the
$(\chi_1,\chi_2)$ rotamer-like sectors are excluded from tuning and
fitting and are used only after the final partition has been fixed.  The CEHS-HM procedure selects
$(\widehat{\lambda},\widehat{K})=(0.525,5)$.

The corresponding mean held-out CEHS-HM score is $0.81545$
(empirical repeated-split SE $0.00370$; $30$ held-out evaluations).
Ten full-data CLARA candidate fits have mean pairwise ARI $0.923$,
minimum $0.855$, and maximum $0.9998$. Accordingly, the five-cluster
representation is used for structural description.

\paragraph{Backbone-side-chain organization.}
The five clusters should be interpreted as joint conformational
modes rather than as estimates of the nine coarse side-chain sectors.
In particular, identical side-chain sectors can split according to
backbone geometry. Among the $4756$ observations assigned to the
$g^{-}/t$ sector, $2757$ occur in Cluster~1 and $1998$ in Cluster~3,
with one additional observation assigned to Cluster~2. These two
major groups correspond to different backbone regimes. Similarly,
among the $1405$ $g^{-}/g^{-}$ observations, $562$ occur in
Cluster~2 and $843$ in Cluster~5. Thus the clustering captures
dependence between backbone and side-chain torsion structure rather
than reproducing a side-chain-only classification.

\paragraph{Post-hoc comparison with ClusTorus.}
For descriptive comparison only, we also rerun the published
ClusTorus ILE workflow on the same observations. This comparison
is not used to tune $\lambda$, choose $K$, or select the CEHS-HM
solution.

The $16$ ClusTorus log-density groups are almost completely nested
within the five broader CEHS-HM modes. Weighted dominant containment
is $99.13\%$, and every ClusTorus group has at least $90\%$ of its
observations in a single CEHS-HM mode. The partition-level
association is
$$
\mathrm{ARI}=0.6722,\quad
\mathrm{NMI}=0.7814,\quad
\mathrm{AMI}=0.7809.
$$
This indicates that the five CEHS-HM clusters provide a coarser
summary of the same conformational organization rather than a direct
replication of the finer ClusTorus partition.

\begin{figure*}[t]
	\centering
	
	\includegraphics[width=0.32\textwidth]
	{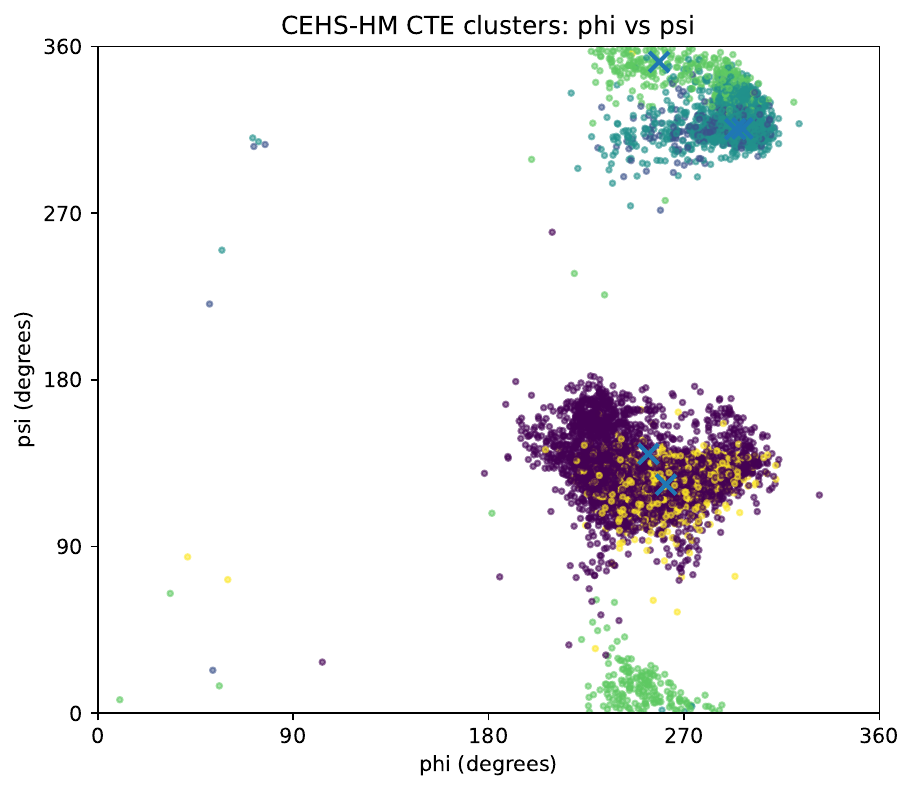}
	\hfill
	\includegraphics[width=0.32\textwidth]
	{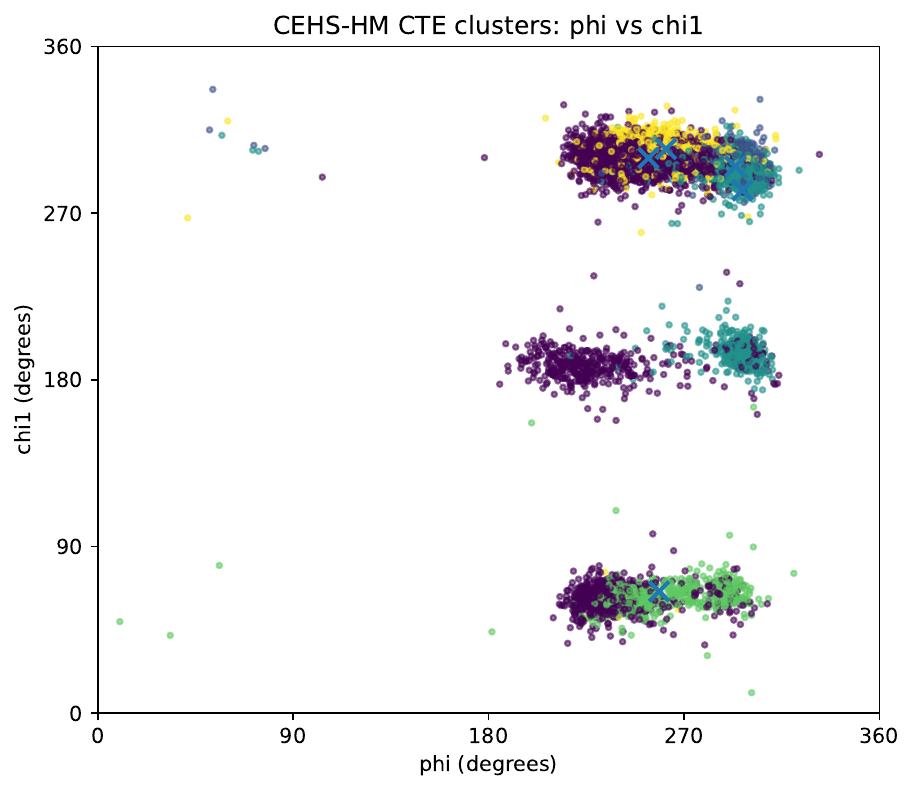}
	\hfill
	\includegraphics[width=0.32\textwidth]
	{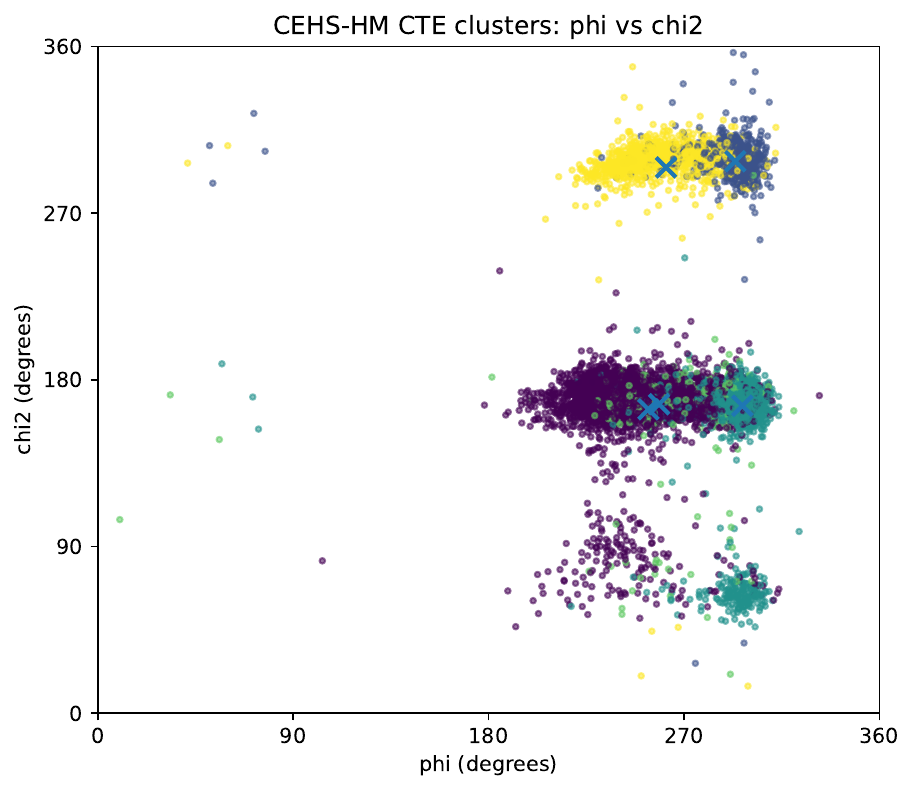}
	
	\vspace{2mm}
	
	\includegraphics[width=0.32\textwidth]
	{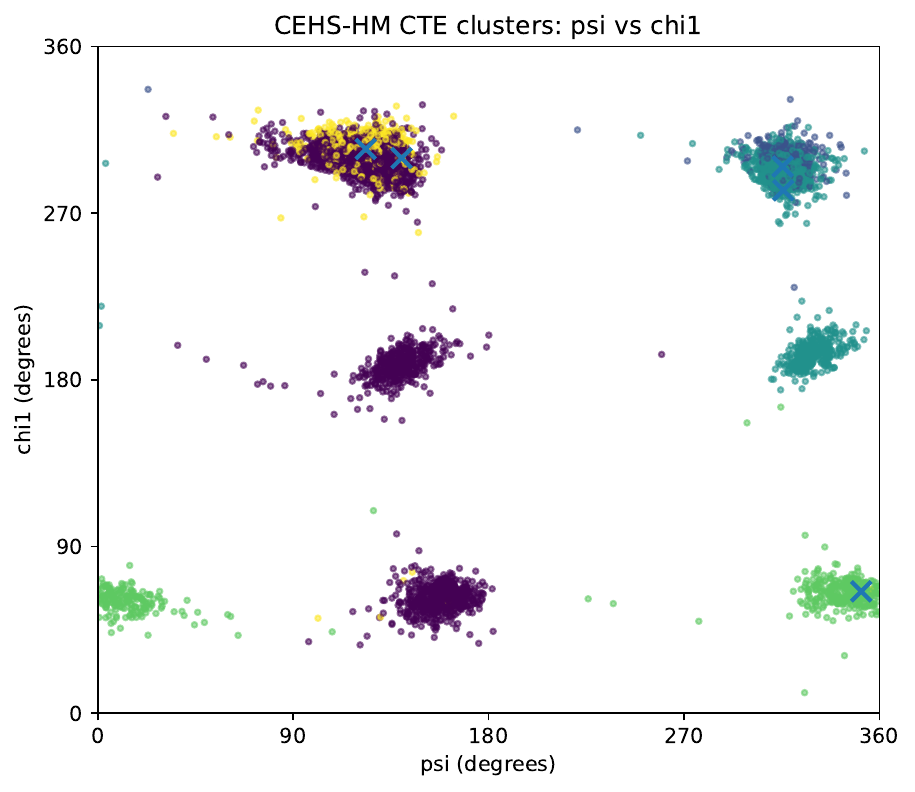}
	\hfill
	\includegraphics[width=0.32\textwidth]
	{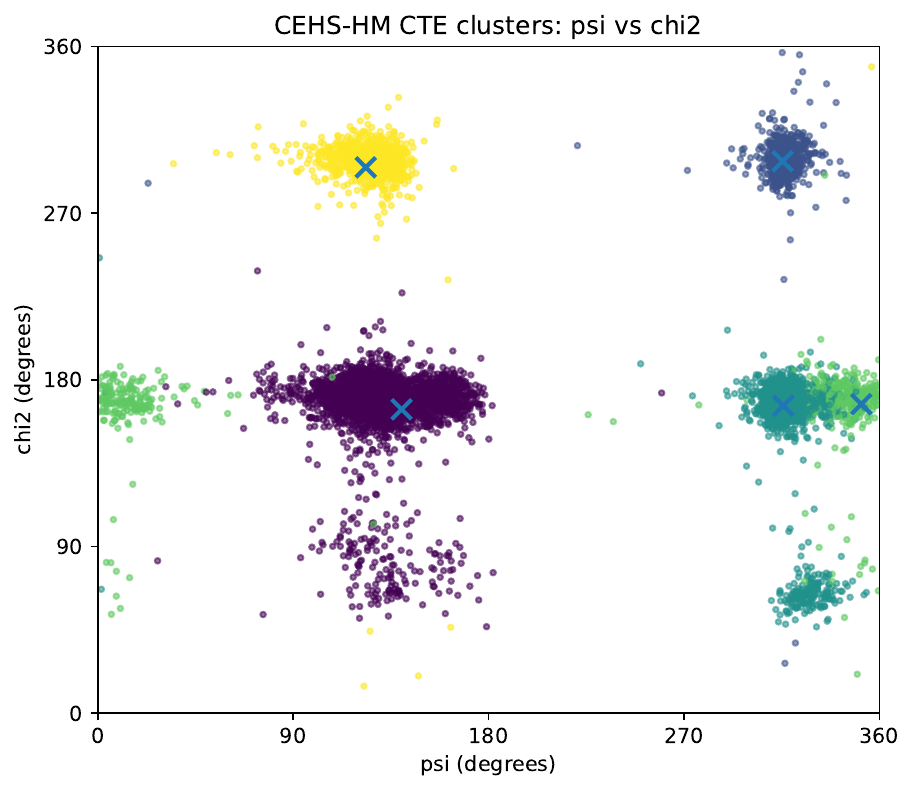}
	\hfill
	\includegraphics[width=0.32\textwidth]
	{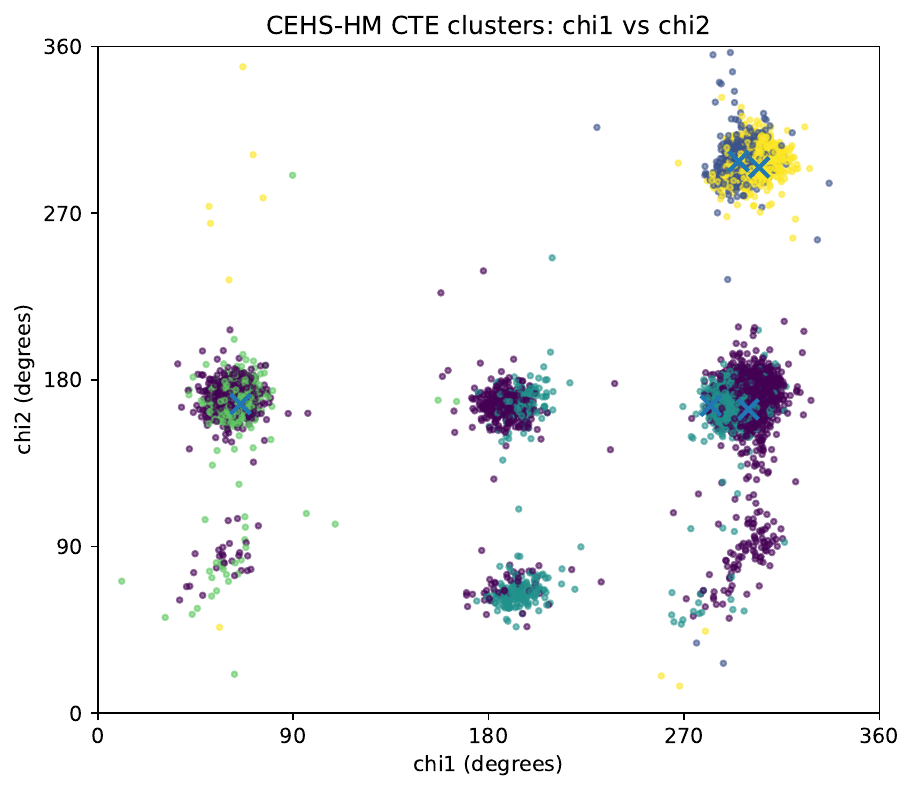}
	
	\caption{All six pairwise angular projections of the
		four-dimensional ILE CEHS-HM partition.
		Colors denote the five clusters selected jointly from the
		four torsion angles; crosses denote the corresponding CTE
		medoids.  The $(\phi,\psi)$ projection displays the backbone
		organization, the $(\chi_1,\chi_2)$ projection displays the
		side-chain organization, and the four cross-projections show
		their joint dependence.  These two-dimensional panels are
		visualizations only: clustering and model selection use the full
		four-dimensional toroidal observations.}
	\label{fig:ile_pairwise}
\end{figure*}
The pairwise projections in
Figure~\ref{fig:ile_pairwise} clarify why the five fitted modes are
not reducible to a side-chain-only classification.  In particular,
the $(\chi_1,\chi_2)$ projection places Clusters~1 and~3 largely in
the same $g^-/t$ region, while the $(\phi,\psi)$ projection separates
them into distinct backbone regimes.  Similarly, Clusters~2 and~5
occupy the same coarse $g^-/g^-$ side-chain sector but differ in
their backbone-angle organization.  The cross-projections further
show that the recovered structure involves joint dependence among
backbone and side-chain torsions.

%\begin{table}[t]
%	\centering
%	\caption{Summary of the five CEHS-HM conformational modes in the
%		ILE data. Angular summaries are circular means in degrees. The
%		dominant rotamer-like sector is used only for post-hoc structural
%		interpretation.}
%	\label{tab:ile_cluster_summary}
%	\small
%	\scalebox{0.75}{\begin{tabular}{rrrrrrrl}
%		\toprule
%		Cluster & $n$ & Fraction &
%		$\bar\phi$ & $\bar\psi$ &
%		$\bar\chi_1$ & $\bar\chi_2$ &
%		Dominant $(\chi_1,\chi_2)$ sector \\
%		\midrule
%		1 & 3844 & 0.476 & -112.0 & 130.5 & -57.4 & 167.6 & $g^-/t$ \\
%		2 &  566 & 0.070 &  -68.2 & -43.3 & -62.8 & -60.4 & $g^-/g^-$ \\
%		3 & 2331 & 0.288 &  -65.4 & -42.6 & -76.3 & 163.1 & $g^-/t$ \\
%		4 &  486 & 0.060 &  -96.3 &  -5.6 &  63.3 & 167.1 & $g^+/t$ \\
%		5 &  853 & 0.106 & -101.2 & 124.8 & -56.3 & -61.3 & $g^-/g^-$ \\
%		\bottomrule
%	\end{tabular}}
%\end{table}
    
\subsection{OpenManipulator-X robotic-arm configurations}
\label{app:omx}

As an application outside structural biology, we analyse the
OpenManipulator-X component of the publicly released Neural Actuation
Dataset (NAD) \citep{dou2026neuralactuator}. For each of the
350 trajectories, we retain one observed representative configuration,
represented by the four revolute-joint angles
$(q_1,q_2,q_3,q_4)\in\mathbb{T}^4$.
Task, subset, condition, and train-validation metadata are excluded
from parameter selection and clustering and are used only for
post-hoc description.
For this analysis, repeated held-out CTE silhouette together with a
one-standard-error parsimony rule selects  $(\widehat{\lambda},\widehat{K})=(0.475,2).$

The unconstrained maximum of the held-out criterion occurs at
$(0.525,2)$, whereas the one-standard-error rule selects the smaller
warping value $0.475$ with essentially comparable predictive
performance.

The two final clusters contain $264$ and $86$ trajectories.
Their circular mean joint angles, in degrees, are
$$
\begin{array}{c|rrrr}
    \text{Cluster} & q_1 & q_2 & q_3 & q_4\\
    \hline
    1 & 355.75 & 2.69 & 1.55 & 72.77\\
    2 & 358.56 & 359.88 & 2.19 & 29.14
\end{array}
$$
so the most pronounced difference occurs in $q_4$.

The fitted partition is also stable under repeated subsampling;
the mean pairwise ARI across the repeated fits is approximately
$0.942$.  Condition labels were not used in fitting.  Post hoc, the
mean dominant-cluster share across the $35$ conditions is $0.96$,
the median is $1.0$, and $27$ conditions are entirely assigned to
one of the two broad configurations.  These summaries are interpreted
descriptively rather than as classification accuracy against
ground-truth task labels.
    
%    \begin{figure}[t]
%        \centering
%        \includegraphics[width=0.88\textwidth]{figures/omx_cte_pairplot.pdf}
%        \caption{OpenManipulator-X pair plot. The two regimes are most clearly separated in $q_4$.}
%    \end{figure}

    %    \bibliographystyle{apalike}
    %    \bibliography{references}
\end{document}